\documentclass[12pt]{spieman}  
\usepackage{amsmath,amsfonts,amssymb}
\usepackage{graphicx}
\usepackage{setspace}
\usepackage{tocloft}
\usepackage{enumitem}
\usepackage{booktabs}
\usepackage{comment}
\usepackage{listings}
\usepackage[skip=10pt]{caption} 
\usepackage{subcaption}
\usepackage{tikz} 
\usetikzlibrary{svg.path} 
\usepackage{pgfplots} 
\usepackage{pgfplotstable} 
\usepackage{adjustbox}
\usepackage{tabu, colortbl}
\usepackage{xr}
\usepackage[super]{nth}
\usepackage{lineno}
\usepackage{textcomp}

\definecolor{mygray}{gray}{0.45}
\definecolor{mycorrect}{rgb}{0, 0, 0} 
\definecolor{light_grey}{rgb}{0.6, 0.6, 0.6}
\definecolor{codegreen}{rgb}{0,0.6,0}
\definecolor{codegray}{rgb}{0.5,0.5,0.5}
\definecolor{codepurple}{rgb}{0.58,0,0.82}
\definecolor{backcolour}{rgb}{0.95,0.95,0.92}

\title{
Simulating Dynamic Tumor Contrast Enhancement in Breast MRI using Conditional Generative Adversarial Networks}

\author[a, b, c, *]{Richard Osuala} 
\author[a]{Smriti Joshi}
\author[d]{Apostolia Tsirikoglou}
\author[a, d]{Lidia Garrucho}
\author[e]{Walter H. L. Pinaya}
\author[b, c]{Daniel M. Lang}
\author[b, c, e]{Julia A. Schnabel}
\author[a, f, **]{Oliver Diaz}
\author[a, g, **]{Karim Lekadir}
\affil[a]{Barcelona Artificial Intelligence in Medicine Lab (BCN-AIM), Departament de Matemàtiques i Informàtica, Universitat de Barcelona, Spain}
\affil[b]{Institute of Machine Learning in Biomedical Imaging, Helmholtz Munich, Munich, Germany}
\affil[c]{School of Computation, Information and Technology, Technical University of Munich, Munich, Germany}
\affil[d]{Department of Oncology-Pathology, Karolinska Institutet, Stockholm, Sweden}
\affil[e]{\textcolor{mycorrect}{School of Biomedical Engineering \& Imaging Sciences, }King's College London, London, United Kingdom}
\affil[f]{Computer Vision Center, Universitat Autònoma de Barcelona, Bellaterra, Spain}
\affil[g]{Institució Catalana de Recerca i Estudis Avançats (ICREA), Passeig Lluís Companys 23, Barcelona, Spain}

\cftpagenumbersoff{figure}
\cftpagenumbersoff{table} 
\begin{document} 

\maketitle

\begin{abstract}
\paragraph*{Purpose:}
Deep generative models and synthetic data generation have become essential for advancing computer-assisted diagnosis and treatment.
We explore one such emerging and particularly promising application of deep generative models, namely, the generation of virtual contrast enhancement. This allows to predict and simulate contrast enhancement in breast magnetic resonance imaging (MRI) without physical contrast agent injection, thereby unlocking lesion localization and categorization even in patient populations where the lengthy, costly and invasive process of physical contrast agent injection is contraindicated.

\paragraph*{Approach:}
We define a framework for desirable properties of synthetic data, which leads us to propose the Scaled Aggregate Measure (SAMe) consisting of a balanced set of scaled complementary metrics for generative model training and convergence evaluation. We further adopt a conditional generative adversarial network to translate from non-contrast-enhanced T1-weighted fat-saturated breast MRI slices to their dynamic contrast-enhanced (DCE) counterparts, thus learning to detect, localize, and adequately highlight breast cancer lesions. Next, we extend our model approach to jointly generate multiple DCE-MRI timepoints, enabling the simulation of contrast enhancement across temporal DCE-MRI acquisitions. Additionally, 3D U-Net tumor segmentation models are implemented and trained on combinations of synthetic and real DCE-MRI data to investigate the effect of data augmentation with synthetic DCE-MRI volumes.
%



\paragraph*{Results:}
Conducting four main sets of experiments, (i) the variation across single metrics demonstrated the value of SAMe, and (ii) the quality and potential of virtual contrast injection for tumor detection and localization were shown. Segmentation models (iii) augmented with synthetic DCE-MRI data were more robust in the presence of domain shifts between pre-contrast and DCE-MRI domains. The joint synthesis approach of multi-sequence DCE-MRI (iv) resulted in temporally coherent synthetic DCE-MRI sequences and indicated the generative model's capability of learning complex contrast enhancement patterns.



\paragraph*{Conclusion:}
Virtual contrast injection can result in accurate synthetic DCE-MRI images, potentially enhancing breast cancer diagnosis and treatment protocols. We demonstrate that detecting, localizing, and segmenting tumors using synthetic DCE-MRI is feasible and promising, particularly considering patients where contrast agent injection is risky or contraindicated. Jointly generating multiple subsequent DCE-MRI sequences can increase image quality and unlock clinical applications assessing tumor characteristics related to its response to contrast media injection as a pillar for personalized treatment planning.

\end{abstract}


{\noindent \footnotesize\textbf{*}Corresponding author: Richard Osuala, \linkable{Richard.Osuala@ub.edu}

\noindent \footnotesize\textbf{**}These authors contributed equally to this work.
}

\begin{spacing}{1} 

\section{Introduction} \label{sect:intro}

\paragraph{Deep learning progress in breast cancer imaging}
Breast cancer was the most common cancer diagnosis worldwide in 2020, taking people of all ages and genders into account. The staggering number of 2.26 million new cases and 684,996 reported deaths underline the significant global burden of breast cancer \cite{globalCancerObservatory}. In breast cancer imaging and medical imaging at large, deep learning has been gaining popularity due to its promising capabilities of sifting through image data to uncover hidden associations. This capacity allows trained deep learning models to recognize subtle patterns in unseen data, which enables solving a plethora of clinical tasks with high potential to improve patient care. With the emergence of deep learning, vast progress has been observed, e.g., in the promising development of automatic methods for the screening, diagnosis, treatment, and monitoring of cancer based on dynamic contrast enhanced magnetic resonance imaging (DCE-MRI). Such methods include the automated tumor detection, localization, segmentation, and characterization for preoperative planning, patient survival assessment, quantification of recurrence risk, and estimation of treatment response \cite{ground_truth, joshi2024leveraging, osuala2022data, lang2023multispectral, radhakrishna2018role, comes2021early}. 

\paragraph{Usage and benefit of contrast agents}
Through the alteration of magnetic properties of tissue, intravenously administered contrast agents (CA), which are commonly based on gadolinium (Gd), manifest as hyper-intense in DCE-MRI. Thus, they allow the visualization of blood flow and changes in permeability. Multiple DCE-MRI volumes are consecutively acquired (before, during, and after CA administration) to enable the time-dependent evaluation of tissue characteristics and assessment of potential abnormalities. The time-signal intensity curves of such dynamic contrast sequences reflect signal intensity changes induced by the uptake and wash-out of CA over time \cite{wu2016breast, el2009dynamic, kuhl1999dynamic}. 
DCE-MRI contrast uptake serves as an important biomarker in oncology, enabling cancer detection, characterization, subtype determination, differentiation of malignancy, recurrence prediction, and treatment response assessment \cite{ground_truth,tao2016dce,chang2017delineation}. Notably, kinetic analysis of contrast enhancement in breast DCE-MRI plays a crucial role in lesion characterization, with features such as peak enhancement, time-to-peak, wash-in and wash-out slopes reflecting malignancy risk. Furthermore, DCE-MRI reveals temporal patterns of contrast enhancement that are not only correlated with breast cancer presence but also offer insights into genetic alterations associated with risk of recurrence, response to chemotherapy, and even the underlying molecular subtypes of breast tumors \cite{wu2016breast, el2009dynamic, kuhl1999dynamic, arasu2011can,chang2017delineation}.

\paragraph{Disadvantages of contrast agents}
%
Despite their undeniable value in diagnostic imaging, gadolinium-based contrast agents (GBCAs) involve concerns regarding their safety profile. For instance, GBCA administration has been linked to an increased risk of nephrogenic systemic fibrosis (NSF)\cite{marckmann2006nephrogenic}. A further concern is given by the deposition of residual gadolinium and its potential bioaccumulation within the body with unknown long-term clinical significance \cite{olchowy2017presence,idee2006clinical,kanda2014high, marckmann2006nephrogenic,nguyen2020dentate}. Following a 2016 European Commission request for a review of GBCAs, the European Medicines Agency (EMA) recommended restrictions on specific intravenous linear agents to mitigate potential risks associated with gadolinium deposition \cite{euagency2017}. Apart from that, safety concerns extend beyond deposition and NSF, as they also comprise well-known acute effects, such as physiologic and allergic-like reactions, as well as Symptoms Associated with Gadolinium Exposure (SAGE) for which the causal relationship to GBCA is still unknown \cite{mcdonald2022symptoms}.
Furthermore, the administration process involves various drawbacks, including lengthy protocols and scan times, significant financial costs, and the requirement for intravenous cannulation and injection procedures. The reliance of DCE-MRI on multiple temporal acquisitions further exacerbates the increased costs and extended examination times for patients. Additionally, susceptibility to motion artifacts necessitates meticulous patient cooperation (e.g., breath-holding) and can be a source of discomfort. Collectively, the aforementioned issues contribute to an undue burden on patients, encompassing both inconvenience and potential risks to well-being \cite{nguyen2020dentate,olchowy2017presence}. Additionally, GBCA administration has been causing the emergence of Gadolinium as a contaminant polluting aquatic ecosystems and environments \cite{rogowska2018gadolinium,dekker2024review} including drinking water supplies, where its respective degradation products can further increase the risk of adverse health effects \cite{brunjes2020anthropogenic}. As a consequence, GBCA administration is contraindicated in multiple scenarios, which include patient populations with adverse reactions, pregnancy, kidney malfunctions, missing consent, or high-risk breast cancer screening populations where GCBA exposure extends recommended thresholds in accumulated dosage or frequency \cite{zhang2023synthesis,osuala2024pre,osuala2024towards}.

\begin{figure}[tb]
\includegraphics[width=\linewidth]{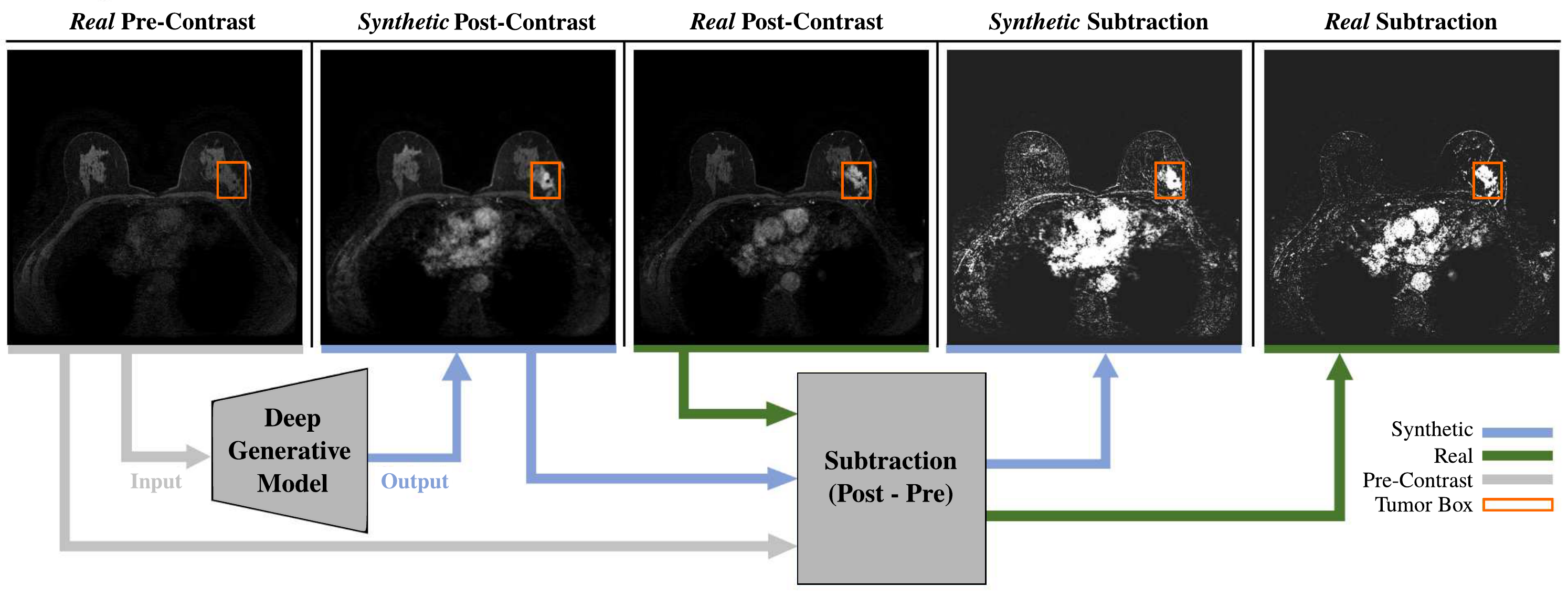}
\centering
\caption[]{Overview of pre- to post-contrast DCE-MRI synthesis using deep generative models, thereby localizing the contrast-enhanced tumor. Extending single sequence to multi-sequence DCE-MRI image generation further allows the characterization of tumors based on their temporal patterns of contrast agent uptake. The resulting synthetic images can be added as training data for downstream tasks (e.g., tumor segmentation), but, as shown, they can also be utilized to compute subtraction images commonly used in clinical settings for the diagnosis and treatment of breast cancer. 
}
\label{fig:synthesis}
\end{figure}

\paragraph{Contribution}

The aforementioned issues emphasize the necessity of alternatives that can be safely issued while also simulating GBCA administration in a way that still yields some of its benefits. To this end, we propose the synthetic generation of DCE-MRI using deep generative models, which constitutes a faster, motion artifact-free, and non-invasive alternative with improved cost-effectiveness that also avoids burdening patient health and well-being. Extending over \textit{Osuala et al. (2024)} \cite{osuala2024pre}, we provide a principled definition of trustworthy synthetic data and introduce the simultaneous generation of images from multiple DCE-MRI timepoints. We further include a quantitative and qualitative evaluation for jointly generated several DCE-MRI timepoints. Additionally, the temporal contrast enhancement is analyzed at the lesion level including a quantification of contrast intensity patterns per patient case and accumulated over the entire dataset. Overall, our work presents the following contributions and novelties to advance the field of synthetic DCE-MRI for breast cancer applications:\vspace{-0.13cm}
\begin{itemize}
    \item \textit{Pre-contrast to DCE-MRI synthesis:}  We implement and validate a conditional generative adversarial image synthesis model capable of translating pre-contrast to DCE breast MRI axial slices, which includes lesion detection, localization, and realistic contrast manifestation.\vspace{-0.13cm}

    \item \textit{Generative model selection framework:} We provide a principled definition of trustworthy synthetic data upon which we derive the Scaled Aggregate Measure (SAMe) and validate it by finding the optimal synthetic data generator. SAMe combines perceptual and pixel-level synthetic data evaluation, enabling comparisons across generative models and the selection of optimal training checkpoints.\vspace{-0.13cm}

    \item \textit{Clinical utility validation of synthetic DCE-MRI volumes for tumor segmentation:}  We demonstrate the potential of our synthetic data by incorporating it into breast tumor segmentation pipelines. This approach enhances robustness across data domains by providing a wider range of training data while showing the coherence of our synthetic axial slices when stacked as 3D MRI volumes.\vspace{-0.13cm}

    \item \textit{Joint synthesis of multiple DCE-MRI timepoints:}  We introduce and empirically validate the joint generation of images from multiple DCE-MRI timepoints using GANs. Further, intensity distributions are extracted from the region-of-interest to model and assess contrast enhancement patterns of real and synthetic data on individual and dataset level.\vspace{-0.13cm} 
\end{itemize}

\section{Related Work} \label{sec:background}

Generative models such as generative adversarial networks (GANs) \cite{goodfellow2014generative} and denoising probabilistic diffusion models (DDPMs) \cite{sohl2015deep,song2019generative,ho2020denoising} and Latent Diffusion Models (LDMs) \cite{rombach2022high} have been widely applied to medical imaging in general and breast imaging in particular \cite{osuala2022data,pinaya2023generative,osuala2023medigan,pasquini2022synthetic}. For example, \textit{Khader et al. (2023)} \cite{khader2023denoising} use unconditional DDPMs to generate non-fat saturated T1-weighted breast DCE-MRI sequences.
A set of models have been proposed to condition the generation process on input images \cite{isola2017image, zhu2017unpaired, choi2018stargan, park2019semantic, sushko2020you, saharia2022palette, zhang2023adding} unlocking image-to-image translation and domain-adaptation applications in medical imaging \cite{wolterink2017deep,garrucho2023high,garcia2024breast}. \textit{Konz et al. (2024)} \cite{konz2024anatomically}, for instance, conditioned LDMs on anatomical segmentation masks to generate pre-contrast breast MRI based on 100 patient cases from the Duke-Breast-Cancer-MRI Dataset \cite{saha2018machine}.

Furthermore, a few first studies started to condition generative models on pre-contrast images to generate their post-contrast counterparts \cite{muller2023using,zhang2023synthesis,osuala2024towards,xue2022bi,chen2022synthesizing}.
For instance, \textit{Kim et al. (2022)} \cite{kim2022tumor} designed a tumor-attentive segmentation-guided GAN that synthesizes a contrast-enhanced T1 breast MRI image from a pre-contrast image, while being guided by the predictions of a surrogate segmentation network. However, with the objective of improving segmentation using GAN-generated data, it can become counterproductive to limit the GAN contrast translation to the tumor segmentation predicted by the segmentation model. Similarly, but based on a chained tumor detection model instead of a segmentation model, \textit{Zhao et al. (2020)} \cite{zhao2020tripartite} introduced Tripartite-GAN to generate contrast-enhanced from non contrast-enhanced liver MRI. As high-quality annotations such as segmentation masks or region-of-interest bounding boxes are costly to annotate and, therefore, likely a scarce resource \cite{osuala2022data}, it is desirable to accomplish the task of pre- to post-contrast synthesis without relying on such annotations. \textit{Wang et al. (2021)} \cite{wang2021synthesizing} introduced a two-stage GAN that, in its first stage segments the contrast enhancement of the T1-weighted image based on an adversarial loss. Next, in its second stage, it is trained to generate post-contrast DCE images relying on the segmentation network from the first stage using an L1-loss, an adversarial loss, and an edge detector based L2 loss.
\textit{Xue et al. (2022)} \cite{xue2022bi} presented a pre- to post-contrast and post- to pre-contrast GAN for brain MRI images. Their bi-directional GAN encodes contrast and image in separate latent representations with the contrast representation producing a contrast enhancement map as output, which can then be subtracted from the synthetic post-contrast image to create the corresponding pre-contrast image.

\textcolor{mycorrect}{Recent work has further demonstrated the potential of virtual contrast agents across various medical imaging modalities \cite{hasny2024myocardial, bone2021contrast, liu2024trustworthy, dar2019image, gong2018deep, cheng2024pixelwise, rofena2024deep, liu2022virtual}. For instance, a feasibility study explored replacing gadolinium with a Bayesian deep learning model that predicts virtual contrast enhancement from non-contrast multiparametric brain MRI \cite{kleesiek2019can}. Similarly, a multicenter, retrospective neuro-oncology study showed that convolutional neural networks can generate synthetic post-contrast T1-weighted MRI from pre-contrast scans, enabling accurate tumor burden assessment comparable to real contrast-enhanced images \cite{preetha2021deep}.}

\textcolor{mycorrect}{In parallel, novel methodological advances have emerged. One example is the Conditional Autoregressive Vision Model (CAVM), which synthesizes contrast-enhanced brain MRI using masked self-attention in an autoregressive framework that simulates progressive contrast agent dosage \cite{gui2024cavm}. Another approach leverages conditional diffusion and flow-matching models to incorporate uncertainty estimation in virtual contrast synthesis \cite{piening2024conditional}.}

\textcolor{mycorrect}{Further, diffusion models have been shown to harmonize varying contrast levels while maintaining consistent segmentation performance in brain MRI \cite{durrer2023diffusion}. Conditional GANs have also been trained on scans acquired at varying GBCA doses to synthesize contrast-enhanced images at fractional levels, using a patch-based Wasserstein loss to preserve noise characteristics \cite{arjovsky2017wasserstein, pinetz2023faithful}. Other works have introduced novel loss functions to improve synthesis fidelity, such as frequency-domain consistency \cite{lei2024ifgan} and tumor-focused learning objectives \cite{chen2022synthesizing}.}

\textit{Müller-Franzes et al. (2023)} \cite{muller2023using} translated T1 and T2 images to contrast-enhanced breast MRI images using pix2pixHD \cite{wang2018high} and conducted an observer study to test the realism of the synthetic images. Furthermore, \textit{Müller-Franzes et al. (2024)} \cite{muller2024diffusion} compare in a more recent work Diffusion Models and GANs for synthesizing higher-dose DCE-breast MRI subtraction images from their lower-dose counterparts. 
\textit{Han et al. (2023)} \cite{han2023synthesis} model the translation of Diffusion Weighted Imaging (DWI) from breast DCE-MRI volumes as sequence-to-sequence translation task, while \textit{Zhang et al. (2023)} \cite{zhang2023synthesis} designed a GAN to synthesize contrast-enhanced breast MRI from a combination of encoded T1-weighted MRI and DWI images
\cite{wang2021synthesizing, xue2022bi, muller2023using, zhang2023synthesis}. 

However, recent promising approaches have not been used to compute synthetic subtraction images and have not been validated 
on their potential to improve tumor segmentation using synthetic data.
Furthermore, these previous approaches generated images from a single post-contrast sequence rather than generating images from multiple temporal DCE-MRI sequences. The latter remains a largely underexplored research problem. To this end, \textit{Schreiter et al. (2024)} \cite{schreiter2024virtual} tested the simultaneous generation of DCE-MRI timepoints with a U-Net architecture using multiple inputs (T1-weighted, T2-weighted, and diffusion-weighted MRI with multiple b-values). In \textit{Osuala et al. (2024)} \cite{osuala2024towards}, the generation of a single, temporally variable DCE-MRI timepoint was shown by conditioning a latent diffusion model on the time-passed since respective pre-contrast acquisition. However, jointly synthesizing DCE-MRI timepoint can be desirable to ensure coherence across the resulting images. While extending over U-Net architectures by adopting a multi-scale conditional GAN, we further note that some modalities are not readily available in all clinical settings (e.g., high-risk population DCE-MRI breast cancer screening without T2-weighted and diffusion-weighted MRI), prompting us to input only single T1-weighted MRI images into our model for contrast synthesis.

By addressing the temporal dynamics in DCE-MRI generation,  we enable a more nuanced radiologic analysis of tumor localization and contrast enhancement patterns required in clinical settings. For instance, we show that our approach is promising to achieve higher image quality and detection sensitivity. At the same time, it also enables the assessment of contrast kinetics, which comprises important clinical biomarkers for cancer characterization such as contrast wash-in and wash-out slopes, peak enhancement, and perfusion and permeability parameters \cite{tao2016dce, ground_truth,wu2016breast, el2009dynamic, kuhl1999dynamic}. 

\section{Materials and Methods}

\subsection{Dataset}
The Duke-Breast-Cancer-MRI Dataset \cite{saha2018machine}, a single-institutional open-access dataset available on The Cancer Imaging Archive \cite{clark2013cancer}, is used in this study. The dataset was acquired at the Duke Hospital in the United States between \nth{1} January 2000 and \nth{23} March 2014. The dataset spans 922 biopsy-confirmed patient cases with invasive breast cancer. It contains information about their histology reports, demographics, treatment records, recurrence and follow-up information, ultrasound and mammography screening information, alongside a set of pre-operative MRI images. Each case involves one fat-saturated T1-weighted sequence (pre-contrast) and up to 4 corresponding fat-saturated T1-weighted DCE sequences (post-contrast). Between each DCE acquisition, a median of 131 seconds passed with scans acquired using a field strength of either 1.5 T or 3 T MRI machine. Out of the 922 patients, 828 were administered either Magnevist\textregistered 
or MultiHance\textregistered 
as CA with a contrast bolus volume ranging from 6mL to 20 mL.
The axial MRI scans come in dimensions of either $320^2$, $448^2$ or $512^2$ in the coronal and sagittal planes, while consisting of a varying number of slices in the axial plane. After transforming the respective DICOM files into 3D NIfTI volumes, their voxel values are min-max normalized per volume and scaled to values in the range $[0, 255]$. 
\textcolor{mycorrect}{
Next, axial slices are extracted from the 3D fat-saturated T1-weighted (DCE-)MRI volumes after resampling them to an isotropic resolution of 1 mm³ using the pixel spacing information provided in the DICOM headers.}

We further source 3D tumor segmentation masks for 254 of our cases from \textit{Caballo et al (2023)} \cite{ground_truth}. \textit{Caballo et al (2023)} segmented these masks automatically using a fuzzy means algorithm in MATLAB. The masks were then refined by an experienced medical physicist and validated by a radiologist. We further manually verified these 254 cases, validating that the masks correctly correspond to the tumor volumes in the first phase of the DCE-MRI acquisition (timepoint 1) and, where necessary, adjusted the orientation.

For the single-sequence pre- to post-contrast synthesis model, we used 668 cases without segmentation masks, out of the total of 922 cases of the dataset, as training data, while the remaining 254 cases with masks are randomly split into validation (224 cases) and test (30 cases) sets. \textcolor{mycorrect}{This split is due to the availability of ground truth tumor masks and avoids data leakage by training the synthesis model exclusively on cases that are not used for validation or test of the segmentation model.} All axial slices, i.e., tumor containing and non-tumor containing slices, were extracted from 3D fat-saturated T1-weighted (DCE-)MRI NIfTI volumes. These slices are then used as train, validation, and test data to enable the model to translate any 2D slice from the 3D volumes.
For the segmentation model, the same test set is used. For training and validation, we exclude 33 multi-focal cases \textcolor{mycorrect}{from the 224 cases where ground truth tumor masks are available}. \textcolor{mycorrect}{Multi-focal cases are defined by containing multiple distinct tumor foci. Their removal allows to ensure consistency with the ground truth annotations which delineate only a single primary tumor lesion.}
We apply a 5-fold cross-validation, splitting the remaining 191 cases into training (80\%) and validation (20\%) subsets. 

For the multi-sequence pre to post-contrast synthesis model, we apply a new random split of the dataset in order to have a higher number of training cases, resulting in \textcolor{mycorrect}{762} train, 50 val, and 100 test cases, \textcolor{mycorrect}{and including only cases with a minimum of 3 available DCE-MRI sequences.} In this multi-sequence scenario, all tumor-containing axial slices were extracted and used alongside an additional 10\% of axial slices adjacent to the tumor (i.e., 5\% before the first and 5\% after the last tumor-containing slice in axial dimension). Tumor containing slices are identified based on the bounding box annotations of the Duke Dataset. Pre- and post-contrast slices are extracted as corresponding pairs.

\subsection{Image Synthesis}

\subsubsection{Generative Adversarial Networks}

GANs \cite{goodfellow2014generative} are a family of deep generative models composed of multi-hidden layer neural networks to implicitly learn a real data distribution from a set of real data samples to then, ultimately, sample unobserved new data points from that distribution.
GANs are based on a two-player min-max game of a generator and a discriminator network. The generator ($G$) strives to create samples ($\hat{x}$) from a noise distribution ($p_{z}$) that the discriminator ($D$) cannot distinguish from samples ($x$) stemming from the real image distribution ($p_{data}$), resulting in the value function:
\begin{equation} \label{eq:1}
\begin{aligned}
\min_{G} \max_{D} V(D,G) = \min_{G} \max_{D}[\mathbb{E}_{x\sim p_{data}} [log D(x)] + \mathbb{E}_{z\sim p_{z}} [log(1 - D(G(z)))]].
\end{aligned}
\end{equation}
%
\textit{Goodfellow et al. (2014)}~\cite{goodfellow2014generative} define the discriminator $D$ as a binary classifier, detecting whether a sample $x$ is either real or generated. The discriminator is trained via binary-cross entropy with the objective of minimizing the adversarial loss function $\mathcal{L}_{adv}$, 
which the generator, on the other hand, tries to maximise:
\begin{equation} \label{eq:2}
\begin{aligned}
\mathcal{L}_{adv} = - \mathbb{E}_{x\sim p_{data}} [log D(x)] + \mathbb{E}_{z\sim p_{z}} [log(1 - D(G(z)))].
\end{aligned}
\end{equation}

\subsubsection{Pre- to Post-Contrast DCE-MRI Synthesis}

\begin{figure} [ht]
\begin{center}
\includegraphics[width=\textwidth]{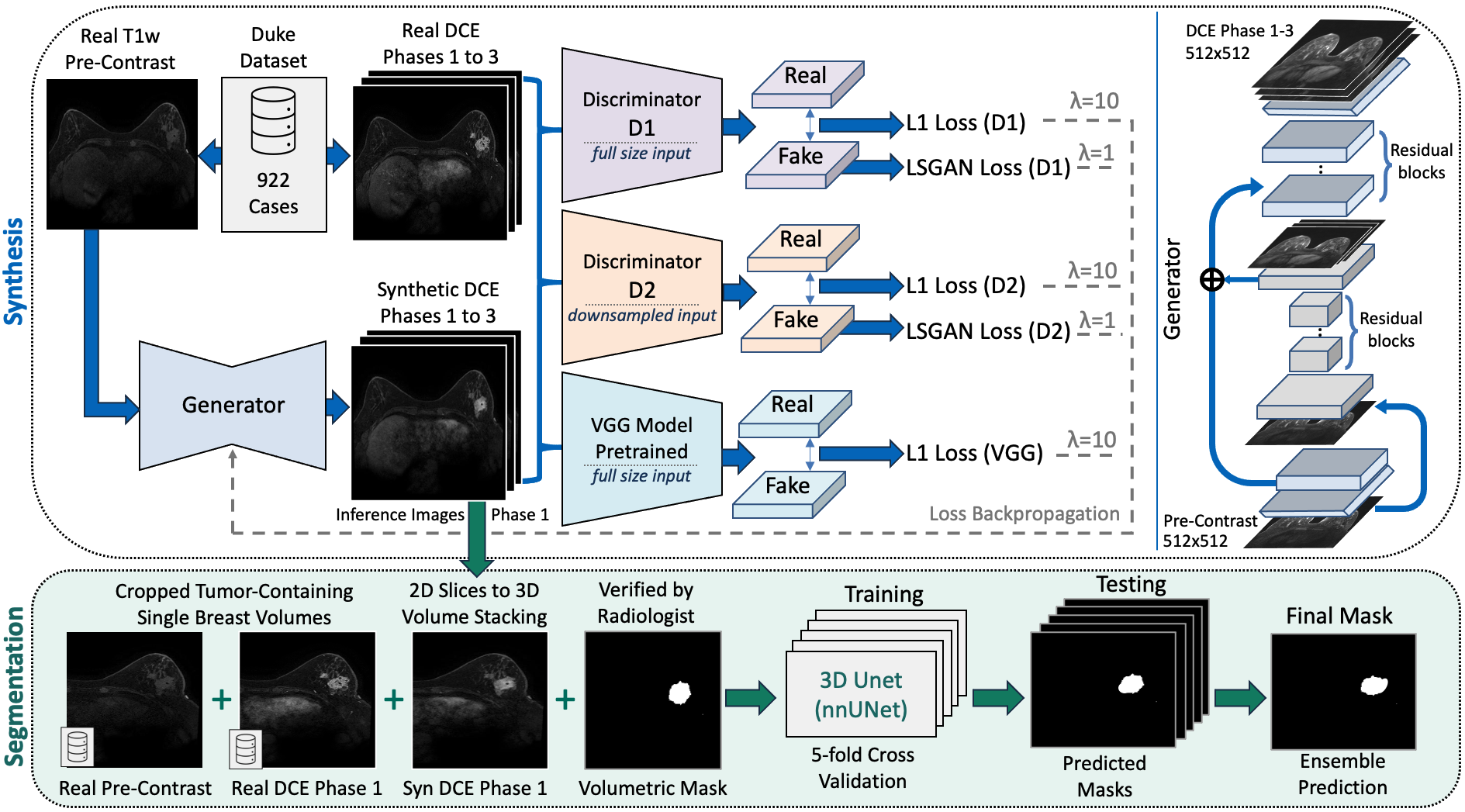}
\end{center}
\caption[] 
{\label{fig:gan_training} Overview of training workflow of our pre- to post-contrast translating GAN based on Pix2PixHD \cite{wang2018high}. Three reconstruction losses (L1) and two least squares adversarial losses (LSGAN)\cite{mao2017least} from two discriminators (D1 \& D2) and one pre-trained VGG \cite{simonyan2014very} model are backpropagated into the generator, where lambda ($\lambda$) represents the weight of each of the different losses. Processing the images at two different scales inside the generator architecture balances local detail and global consistency\cite{wang2018high}, which is further enforced by the two different image input scales in D1 (full size) and D2 (downsampled).
The segmentation method is based on 3D U-Nets \cite{ronneberger2015u} from the nnU-Net \cite{isensee2021nnu} framework. The iteratively translated synthetic post-contrast axial slices are stacked to create 3D breast MRI volumes. These synthetic volumes correspond to the tumor segmentation masks, which were initially acquired based on the real post-contrast fat-saturated sequence.
}
\end{figure}

In the context of image-to-image (I2I) translation, instead of generating data from a noise distribution, GANs \cite{goodfellow2014generative} receive an input sample from a source distribution ($x$) to generate a corresponding output sample from a target distribution ($\hat{y}$). In this research, we implement a Pix2PixHD\cite{wang2018high} GAN for translating pre-contrast to post-contrast images. Pix2PixHD was chosen for its proven effectiveness in producing high-quality cancer imaging data \cite{osuala2022data}, along with its network architecture and methodological approach specifically tailored for paired image-to-image translation, fitting the pre-to-post-contrast transformation scenario.
As illustrated in Figure \ref{fig:gan_training}, the GAN architecture comprises a generator network that processes images at two different scales—one to ensure global consistency and the other to generate fine details. Additionally, the model incorporates two identical discriminator networks, each operating at different image scales based on downsampled versions of the input images. The training of the model involves a weighted combination ($\lambda$) of least squares adversarial losses \cite{mao2017least} ($\lambda_{adv}=1$), discriminator feature matching losses ($\lambda_{fm}=10$) calculated as the summed L1-loss between the real and synthetic image features extracted by each of the two discriminators, and a VGG-based \cite{simonyan2014very} perceptual loss ($\lambda_{per}=10$):
\begin{equation}
\mathcal{L}_{\text{GAN}} = 
\lambda_{adv} \cdot ( \mathcal{L}_{\text{adv(D1)}} +
\mathcal{L}_{\text{adv(D2)}}) +
\lambda_{fm} \cdot ( \mathcal{L}_{\text{fm(D1)}} + 
\mathcal{L}_{\text{fm(D2)}}) +
\lambda_{per} \cdot \mathcal{L}_{\text{per}}.
\end{equation}
Input images are transformed into the range $[-1, 1]$ and have a probability of 50\% of being rotated by 90 degrees during training. \textcolor{mycorrect}{Following the best practices from \textit{Wang et al (2018)} \cite{wang2018high}}, the model is trained for 200 epochs using an Adam optimizer ($\beta=0.5$) \cite{kingma2014adam} and a learning rate of 2e-4 that decays linearly to zero from epoch 100 to 200. The images in the dataset are resized to pixel dimensions of $512 \times 512$.

In the case of (i) pre-contrast to phase 1 DCE-MRI image synthesis, the grayscale post-contrast (phase 1) image is duplicated three times and stacked into 3 channels. This model was trained on a single NVIDIA GeForce RTX 3090 GPU with 24GB RAM using a batch size of 1. 

For the case of (ii) pre-contrast to multi-DCE phase synthesis, the respective images of DCE phase 1, 2, and 3 are concatenated resulting in an output pixel dimension of $512 \times 512 \times 3$. In this latter case, and despite outputting a single image, the pix2pixHD learns to synthesize the first three DCE-MRI acquisitions jointly. We extract the output of each of the three channel dimensions from these output images and store them separately as $512 \times 512 \times 1$ image per DCE phase. We specifically selected the first three DCE-MRI phases, as additional acquisitions, such as DCE phase 4 and phase 5, are not available for a considerable fraction of cases in the dataset. This model was trained on a single Nvidia RTX A6000 GPU with 48GB RAM and a batch size of 8 using the PyTorch library \cite{pytorchPaszke} in a Python 3.11 environment.

\subsection{Synthetic Data Evaluation}

\subsubsection{Defining Trustworthy Synthetic Data} \label{sec:evaluation}

\begin{figure}[h]
\includegraphics[width=\textwidth]{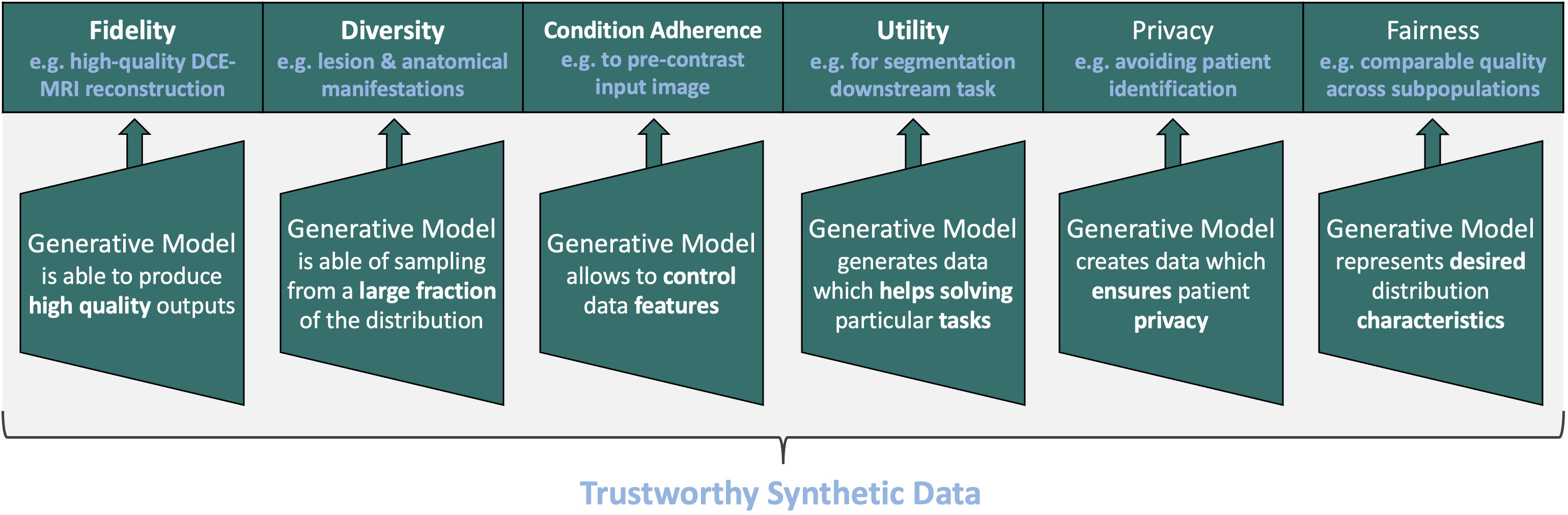}
\caption{Depiction of the generally applicable dimensions of trustworthy synthetic data alongside respective examples (in blue font) for their adoption in deep generative models for medical image synthesis. The present study evaluates fidelity, diversity, condition adherence, and utility. Privacy and fairness are included herein for the comprehensiveness of the dimensions encompassing trustworthy synthetic data.}
\label{fig:synthetic_data_eval}
\end{figure}

Our review of the medical image synthesis literature \cite{osuala2022data,muller2023using, wang2021synthesizing,xue2022bi,kim2022tumor, zhang2023synthesis,zhao2020tripartite, pasquini2022synthetic} described in Section \ref{sec:background} reveals a lack of agreement on the appropriate metrics for assessing synthetic imaging data.
Different metrics provide particular strengths such as correlation with human visual perception or usefulness for clinical application. In order to prioritize and select specific metrics, we note the need for a principled definition of what desirable trustworthy synthetic data should encompass. To this end, inspired by the SynTRUST framework \cite{osuala2022data}, while also building upon insights from previous synthetic data evaluation studies and guidance provided in the literature \cite{osuala2022data,osuala2023medigan,osuala2024enhancing,xing2023you,woodland2023importance,borji2019pros,borji2022pros,osuala2024pre}, we provide a general definition of trustworthy synthetic data.
As depicted in Fig. \ref{fig:synthetic_data_eval}, we define trustworthy synthetic data as multi-dimensional framework comprising synthetic data (a) fidelity, (b) diversity, (c) condition adherence, (d) utility, (e) privacy, and (f) fairness.

Fidelity (a) refers to the quality, realism, and degree to which the generated images accurately and convincingly replicate the characteristics of real-world images. Beyond visual similarity, fidelity also requires preserving essential features, shapes, textures, and statistical properties of the original data.
Diversity (b) represents the objective of generating a wide range of synthetic images, ideally capturing the full spectrum of variability present in real-world images. This includes variations in features such as intensities, texture, structure, and patterns, as well as domain shifts within the data such as differences in pathological and anatomical manifestations, viewpoints, and contexts.
Condition adherence (c) describes the extent to which the generated images accurately reflect specific, predefined conditions or attributes that were set during the synthesis process. This ensures that the images conform to particular requirements or constraints, such as provided features, labels, variables, contextual details, or input images.
Utility (d) represents the practical value and effectiveness of synthetic images, commonly tied to achieving one or multiple specific goals or applications. This includes how actionable synthetic images are in serving their intended purposes, such as training machine learning models, validating algorithms, translating data to another domain, or augmenting real-world datasets. Utility can be measured indirectly by quantifying the contribution and impact of synthetic images in a practical task, e.g., via ablation studies.
%
Privacy (e) pertains to the ability of synthetic images to protect sensitive information to ensure confidentiality while still providing valuable data for analysis or training. On one hand, generated data is desired to maintain essential characteristics of the real-world dataset in order to serve specific applications, however, without revealing or compromising confidential or identifiable information such as for instance a patient's ailments, identifying visual features or other related personal data.
Fairness (f) involves ensuring that synthetic images represent diverse and balanced data across different groups or conditions, avoiding biases that could lead to undesirable outcomes. Examples include generating synthetic images that accurately reflect specific or specifically-balanced demographics, conditions, or scenarios without favoring or underrepresenting any particular subpopulation. Hence, synthetic images are not to create or reinforce biases or inequities, thereby promoting equitable, impartially and inclusive performance on downstream applications.

\subsubsection{Synthetic Data Evaluation Metrics for DCE-MRI Synthesis} 
As discussed above, it is desirable to provide a multi-faceted analysis of synthetic data while also automating the evaluation process to avoid laborious and costly manual human expert observer revisions \cite{borji2019pros, borji2022pros, osuala2022data}. 
We assess our synthetic DCE-MRI using multiple metrics comparing them against their real counterparts hence measuring (a) fidelity, (b) diversity, and (c) condition adherence. To this end, we provide comparisons of real-synthetic image pairs and of real and synthetic image distributions. The latter is evaluated using the Fréchet Inception Distance (FID) \cite{heusel2017gans}, which computes the distance between two sets of features each extracted from one imaging datasets. These features are latent representations generated by passing each image in the dataset through a pretrained deep learning model, e.g. an Inception v3 \cite{szegedy2016rethinking} pretrained on natural images from ImageNet\cite{deng2009imagenet} (FID$_{Img}$) or pretrained on radiology images from RadImageNet\cite{mei2022radimagenet} (FID$_{Rad}$). 
Once the latent features are extracted from both synthetic and real datasets, they are each fitted to a multi-variate Gaussian $X$=real and $Y$=synthetic having means $\mu_{X}$ and $\mu_{Y}$ and covariance matrices $\Sigma_{X}$ and $\Sigma_{Y}$ in order to compute the Fréchet distance (FD) as 
        $\textit{FD}(X, Y) = \lVert \mu_{X} - \mu_{Y} \rVert_{2}^{2} + \text{tr}(\Sigma_{X} + \Sigma_{Y} -2(\Sigma_{X}\Sigma_{Y})^{\frac{1}{2}}).$
\textcolor{mycorrect}{A recently introduced variation is the Fréchet Radiomics Distance \cite{osuala2024towards} (FRD), which compares extracted handcrafted radiomics \cite{lambin2012radiomics,van2017computational} feature distributions extraced from medical images.}
To evaluate corresponding real-synthetic DCE-MRI image pairs, we use a comprehensive set of metrics including the mean squared error (MSE), mean absolute error (MAE), peak signal-to-noise ratio (PSNR), multi-scale (MS) structural similarity index measure (SSIM) \cite{wang2004image}, as well as the Learned Perceptual Image Patch Similarity (LPIPS) \cite{zhang2018unreasonable}. Given the presence of corresponding reference images in the pre- to post-contrast translation of axial MRI slices, we average metrics across all MRI slice image pairs, reporting each metric alongside its standard deviation across image pairs.
Lastly, we additionally assess the generated DCE-MRI images
based on their (d) utility by measuring their impact when included as additional breast tumor segmentation model training data, as described in Section \ref{sec:segmentation}.

\subsubsection{Scaled Aggregate Measure (SAMe)} 
As discussed in Section \ref{sec:background} and Section \ref{sec:evaluation}, there is no consensus on methods and metrics for evaluating of synthetic data in general, and in image-to-image synthesis tasks specifically. Various metrics are employed and reported, but there is ambiguity about which metric should take precedence, particularly when different metrics yield conflicting results. This issue also complicates determining the optimal stopping point for training a generative model. Overall, this evident lack of a consistent evaluation metric underscores the necessity for our proposed unified measure of synthetic data quality. 
Given that each metric captures different facets of truth, we suggest that the most effective way to evaluate synthetic data is through an ensemble of metrics. Therefore, we propose a \textit{Scaled Aggregate Measure} (SAMe) that scales and combines several metrics. These metrics include the SSIM, MSE, MAE, FID$_{Img}$\cite{deng2009imagenet,heusel2017gans}, and FID$_{Rad}$\cite{mei2022radimagenet,osuala2023medigan}.
In this work, for simplicity, we define SAMe based on the aforementioned five metrics, but note that the integration of further metrics, such as LPIPS or PSNR, as well as additionally distinguishing between image-level and lesion-level metrics, can further improve the expressiveness and comprehensiveness of SAMe. 
The metrics in SAMe are scaled to a range $[0, 1]$ using per-metric min-max normalisation to achieve comparability and allow their combination. After scaling, smaller values correspond to increased performance for SAMe and each of its internal metrics, including SSIM which was reversed after scaling (i.e., the smaller, the better). Next, we compute SAMe as a non-weighted average between these metrics. 
The choice of the metrics in SAMe is flexible as we motivate researchers to adopt SAMe based on their particular image synthesis problem at hand. However, we draw attention that our choice of metrics comprises a complementary selection balancing perceptual metrics that capture global semantics of imaging features (FIDs), metrics of perceived quality of images (SSIM and FIDs) and metrics focusing on fine-grained pixel-level comparisons between generated and target images (MAE and MSE) to assess the accurateness of replication between an image pair. While FID is associated to a high sensitivity to small changes and high correspondence to human inspection~\cite{borji2022pros}, the pixel level metrics in SAMe measure objective (MSE, MAE) and perceived image fidelity based on a combination of luminance, contrast, and structural information (SSIM)~\cite{samajdar2015}. We further combine analytical metrics (SSIM, MAE, MSE) with metrics derived from latent space feature representations of neural networks (FIDs), with the latter being further divided into domain-agnostic (FID$_{Img}$) and radiology domain-specific (FID$_{Rad}$) features to capture different dimensions of relevant information within the synthetic data. To this end, we compress complementary and mutually exclusive information present in the selected image quality metrics into a single meaningful measure and show its application for the problem of generative model training stopping criterion definition and training checkpoint selection.

\subsection{Tumor Segmentation Downstream Task} \label{sec:segmentation}

The segmentation of tumors is an important clinical task used, among others, to analyze and quantify the tumor and its volume. An accurate tumor delineation can be used for surgery and radiation treatment planning as well as monitoring evaluating tumor growth or decline e.g. before, during and after neoadjuvant chemotherapy.

To evaluate synthetic data for automated tumor segmentation, we adopt a single 3D U-Net\cite{ronneberger2015u} model using the nnU-Net framework \cite{isensee2021nnu} (i.e., \textit{nnunetv2 3d full\_res}). nnU-Net is a versatile deep learning framework for medical image segmentation, which self-configures its architecture based on the input data. We adopt its 3D U-Net architecture to capture volumetric tumor context while also retaining fine details via its skip connections between encoding and decoding layers. Using only the 3D convolutional architecture variant of the nnU-Net framework further enables testing whether individually translated synthetic 2D breast MRI slices can be reassembled to useful 3D breast MRI volumes. We follow the vanilla nnU-Net training configuration, however, without applying any of nnU-Net's post-processing techniques such as all-but-largest-component suppression \cite{isensee2021nnu}, which are not specifically targeted to our breast tumor segmentation task. We train one 3D U-Net tumor volume segmentation model for 500 epochs for each fold in a 5-fold cross-validation (CV). Test set performance is measured based on the averaged predictions of the ensemble of the five models, each of which was trained during one of the CV folds. The Dice coefficient is used as training loss and test set evaluation metric to measure segmentation performance. The Dice coefficient quantifies the overlap between the predicted tumor segmentation $(A)$
and the ground truth $(B)$ in a range [0, 1], with 0 representing no overlap and 1 indicating complete overlap. 

As shown in Fig. \ref{fig:gan_training}, the 2D synthetic slices are stacked to 3D synthetic volumes before being integrated as additional training data into our tumor segmentation pipeline. The same segmentation masks, which had initially been annotated in (a) the first real post-contrast DCE-MRI sequence, are used as labels in the segmentation model for (a) real post-contrast, (b) real pre-contrast, and (c) synthetic post-contrast MRI volumes, as depicted in Figure \ref{fig:image_comparison}.
Given that in the ground truth masks only the primary lesion was annotated, we remove multifocal cases (33 cases) from the segmentation dataset. We further crop the volumes to include only a single breast per image rather than both breast to avoid any bilateral cases and apply Bias field correction \cite{tustison2010n4itk}. The segmentation models were trained on a single NVIDIA GeForce RTX 3090 GPU with 24GB RAM.

\section{Experiments and Results} \label{sec:results}

\subsection{Selection of Generative Model}

\begin{figure}[h]
    \begin{subfigure}[b]{0.565\textwidth}
        \centering
        \includegraphics[width=\textwidth]{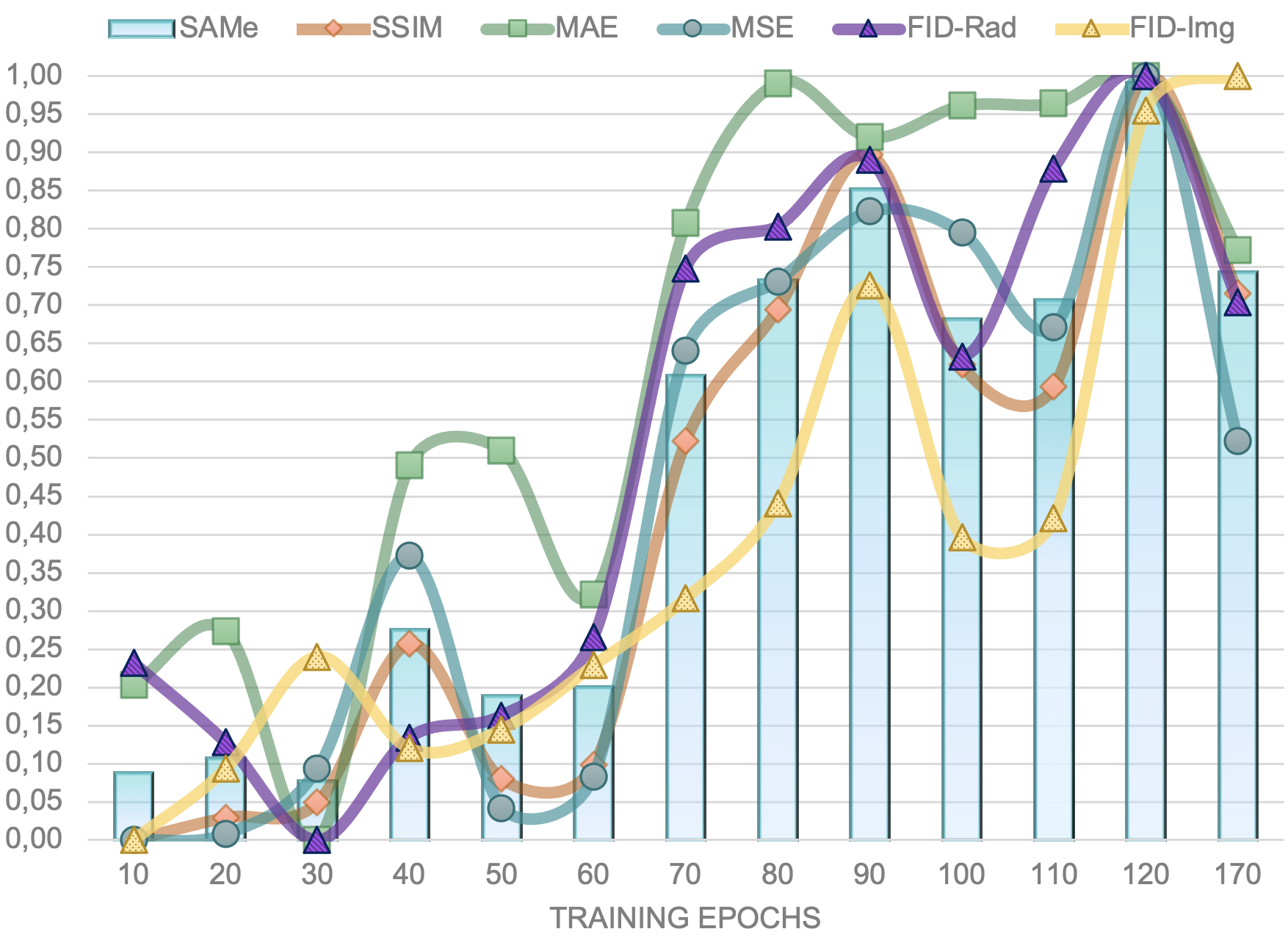}
        \caption{We inspect overall synthetic data fidelity, and diversity using the Scaled Aggregate Measure (SAMe) across generative model training epochs, thereby enabling an informed selection of the best training checkpoint (i.e., epoch 30, achieving the lowest SAMe). Our SAMe metrics include synthetic image distribution distances (FID$_{Img}$ and FID$_{Rad}$), pixel space objective (MSE and MAE) and perception-based (SSIM) quality metrics. Metrics are scaled in range [0,1], where lower values indicated better performance. 
        }
        \label{fig:same_chart}
    \end{subfigure} 
    \hspace{0.1cm}
    \begin{subfigure}[b]{0.41\textwidth}
        \centering
        \includegraphics[width=\textwidth]{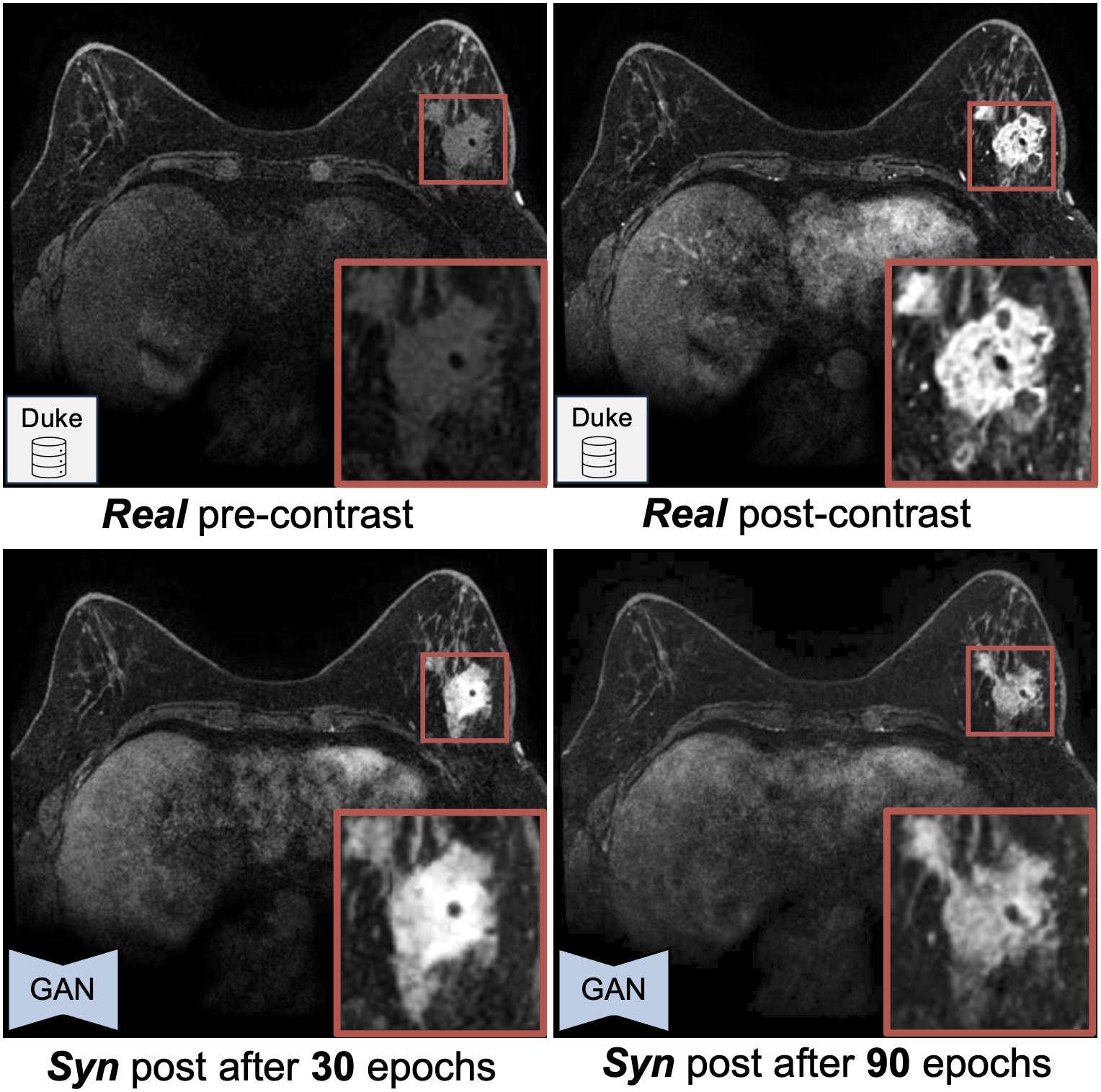}
        \caption{An illustration of synthetic breast DCE-MRI images generated during GAN training after epochs 30 and 90, exhibiting discernible variances in tumor representation. From left to right, (i) the real pre-contrast image is shown (i.e., the GAN input image), (ii) the respective real DCE phase 1 image, the synthetic real DCE phase 1 counterpart after (iii) 30 and (iv) 90 training epochs.}
        \label{fig:same_example}
    \end{subfigure}
\caption{Quantitative (a) and qualitative (b) illustrations of SAMe applied to DCE-MRI synthesis}
\label{fig:same}      
\end{figure}


As the first experiment, we apply our proposed SAMe to our single sequence DCE-MRI generative model, which translates a T1-weighted pre-contrast image to its first sequence (phase 1) DCE-MRI counterpart. To this end, we demonstrate SAMe's effectiveness as a generative model weight selection criterion. As shown in Fig. \ref{fig:same_chart}, we compute SAMe and its internal metrics (FID$_{Img}$, FID$_{Rad}$, MAE, MSE, SSIM) during generative model training on each \nth{10} epoch up until epoch 170. The FID metrics are computed for 3000 and MSE, MAE and SSIM metrics for 5000 synthetic-real post-contrast axial MRI slice pairs from the validation set. 
In Fig. \ref{fig:same_chart}, epochs with metric values close to 0 indicate good generative model performance in comparison to values close to 1. For completeness, we also provide the original values of the metrics before SAMe scaling in Table \ref{tab:image_quality_VAL}. SAMe, as the aggregate across complementary metrics, is depicted using bar charts (in blue) in Fig. \ref{fig:same_chart}. Hence, the shortest bar represents the generative model checkpoint which performs the best across metrics and epochs. We note that the generative model achieves good performance already early in training with a SAMe score of 0.087 in epoch 10. The only checkpoint achieving a better results is the one in epoch 30, resulting in a SAMe of 0.077. Further training follows a trend of gradually reduced performance in SAMe (e.g., 0.682 in epoch 100) likely indicating an increasing overfitting on the training dataset. 
In the last row of Table \ref{tab:image_quality_VAL} real post-contrast images are compared to their real pre-contrast counterparts. Compared to this baseline, the synthetic images from the different epochs (e.g., ep10, ep30, ep50) consistently result in overall better metrics (e.g., FID$_{Img}$, SSIM, MSE, albeit with the exception of MAE) when compared to real post-contrast images from the validation set. 

A noteworthy observation is that, both across and within training epochs, the various metrics yield inconsistent conclusions regarding the optimal synthesis model, highlighting the need for a unified measure such as SAMe. Yet, all metrics exhibit a similar overall trend, with lower (better) values until epoch 60, after which they increase remarkably, suggesting overfitting and diminishing returns from continued training. based on SAMe, epoch 30 emerges as the optimal source for a model checkpoint for generating synthetic post-contrast samples, which are subsequently used for the tumor segmentation downstream task and image synthesis test set evaluation. 

\begin{table}[htb]
\centering
\caption{Example of quantitative image quality results, based on SAMe and reported with standard deviation on the validation set where applicable, for different GAN training epochs (\emph{ep}). FID$_{Img}$, FID$_{Rad}$ are based on 3000 and MSE, MAE, SSIM are based on 5000 synthetic-real DCE-MRI axial MRI image slice pairs. As upper bound, the \emph{Real DCE vs. Real Pre} baseline compares corresponding real pre-contrast and real DCE pairs. 
}
\vspace{2mm}
\resizebox{1.0\columnwidth}{!}{
\begin{tabular}{lccccccc}
    \toprule
     & & \multicolumn{6}{c}{Metric} \\
     \arrayrulecolor{light_grey} \cmidrule(lr){3-8}
    Comparison & Dataset &  FID$_{Img}$ $\downarrow$ & FID$_{Rad}$ $\downarrow$ & SSIM $\uparrow$ & MAE $\downarrow$ & MSE $\downarrow$ & SAMe $\downarrow$ \\
    \arrayrulecolor{black} \toprule
    Real DCE vs. Syn$_{ep10}$ DCE & Val & \textbf{15.047} & 0.108 & \textbf{0.701 $\pm$ 0.081} & 93.895 $\pm$ 41.748 & \textbf{37.803 $\pm$ 9.960} & 0.087 \\
    Real DCE vs. Syn$_{ep30}$ DCE    & Val & 17.308 & \textbf{0.081} & 0.699 $\pm$ 0.081 & 88.733 $\pm$ 39.426 & 38.334 $\pm$ 9.582 & \textbf{0.077} \\
    Real DCE vs. Syn$_{ep50}$ DCE    & Val & 16.412 & 0.089 & 0.696 $\pm$ 0.090 & 101.696 $\pm$ 44.672 & 38.045 $\pm$ 10.985 & 0.188 \\
    Real DCE vs. Syn$_{ep100}$ DCE   & Val & 18.778 & 0.219 & 0.669 $\pm$ 0.116 & 113.144 $\pm$ 59.360 & 42.320 $\pm$ 17.792 & 0.682 \\
    Real DCE vs. Real Pre & Val & 34.062 & 0.120 & 0.660 $\pm$ 0.090 & \textbf{66.146 $\pm$ 31.758} & 42.933 $\pm$ 11.528& \\
    
    %
    \arrayrulecolor{black} \bottomrule
\end{tabular}
}
\label{tab:image_quality_VAL}
\end{table}
%

\subsection{Synthesis of First DCE-MRI Sequence} \label{sec:single-dce-mri-exp}

\begin{figure}
    \centering
    \begin{tikzpicture}
    \draw (0, 0) node[inner sep=0] {
    \includegraphics[height=1.4cm, width=.164\textwidth, trim={1.5cm 6.5cm 1.5cm 2.5cm},clip]{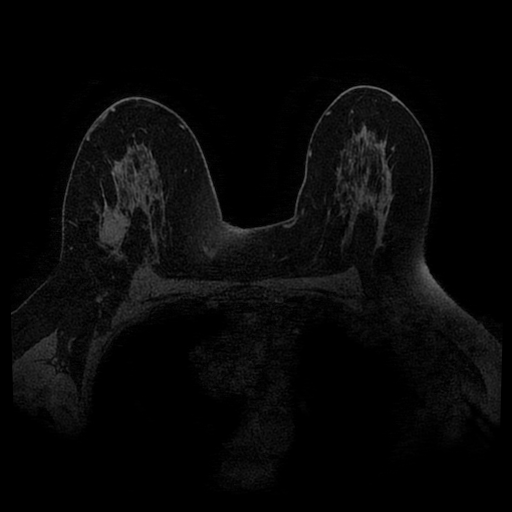}
            };
    \draw (0, 1.0) node {\tiny(\emph{a}) pre-contrast MRI};
    \end{tikzpicture}\hspace{0cm}%
    \begin{tikzpicture}
    \draw (0, 0) node[inner sep=0] {
    \includegraphics[height=1.4cm, width=.164\textwidth, trim={1.5cm 6.5cm 1.5cm 2.5cm},clip]{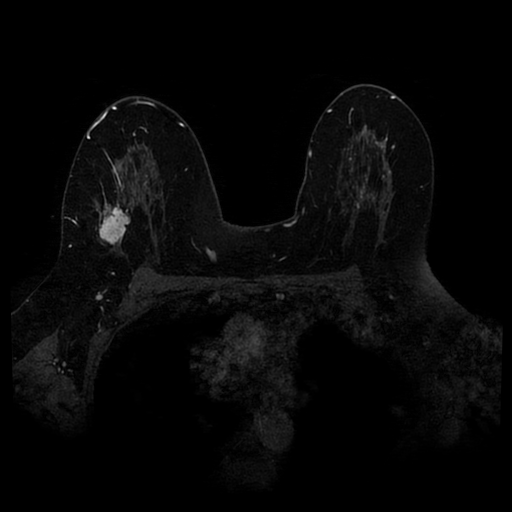}
        };
    \draw (0, 1.0) node {\tiny(\emph{b}) real DCE-MRI};
    \end{tikzpicture}\hspace{0cm}%
    \begin{tikzpicture}
    \draw (0, 0) node[inner sep=0] {
    \includegraphics[height=1.4cm, width=.164\textwidth, trim={1.5cm 6.5cm 1.5cm 2.5cm},clip]{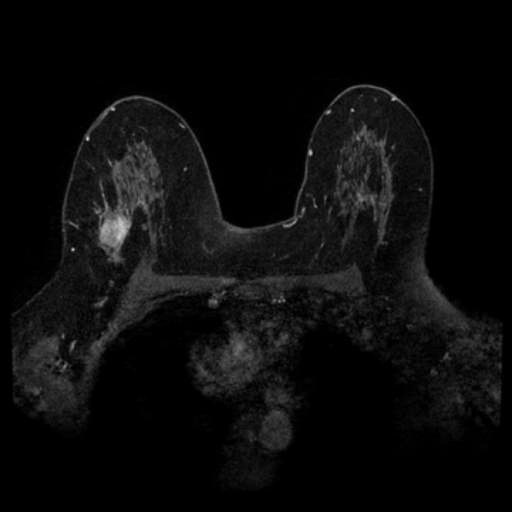}
    };
    \draw (0, 1.0) node {\tiny(\emph{c}) syn DCE-MRI};
    \end{tikzpicture}\hspace{0cm}%
    \begin{tikzpicture}
    \draw (0, 0) node[inner sep=0] {
    \includegraphics[height=1.4cm, width=.164\textwidth, trim={1.5cm 6.5cm 1.5cm 2.5cm},clip]{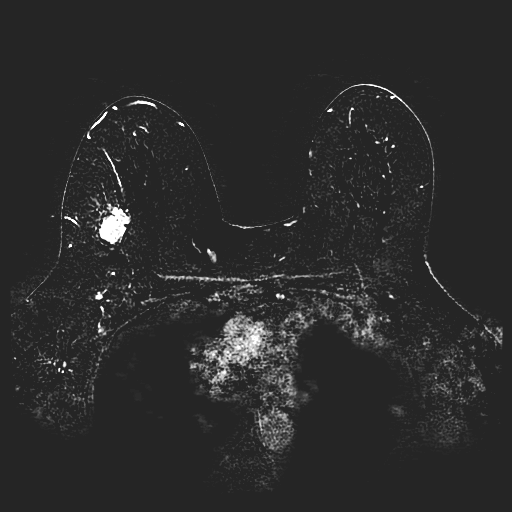}
    };
    \draw (0, 1.0) node {\tiny(\emph{d}) real subtraction};
    \end{tikzpicture}\hspace{0cm}%
    \begin{tikzpicture}
    \draw (0, 0) node[inner sep=0] {
    \includegraphics[height=1.4cm, width=.164\textwidth, trim={1.5cm 6.5cm 1.5cm 2.5cm},clip]{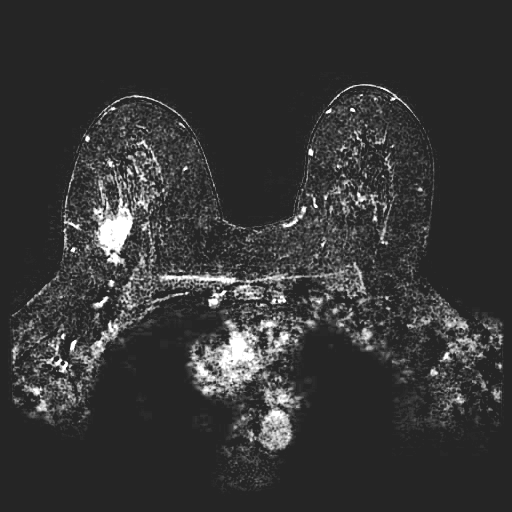}
    };
    \draw (0, 1.0) node {\tiny(\emph{e}) syn subtraction};
    \end{tikzpicture}\hspace{0cm}%
    \begin{tikzpicture}
    \draw (0, 0) node[inner sep=0] {
    \includegraphics[height=1.4cm, width=.164\textwidth, trim={1.5cm 6.5cm 1.5cm 2.5cm},clip]{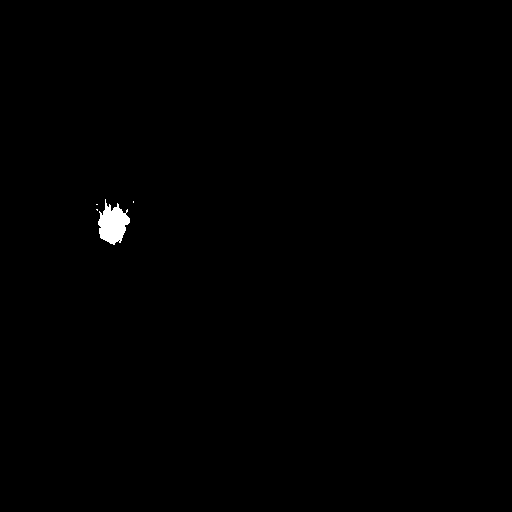}
    };
    \draw (0, 1.0) node {\tiny(\emph{f}) ground truth mask};
    \end{tikzpicture}\hspace{0cm}%


    
    \includegraphics[height=1.4cm, width=.164\textwidth, trim={1.5cm 6.5cm 1.5cm 2.5cm},clip]{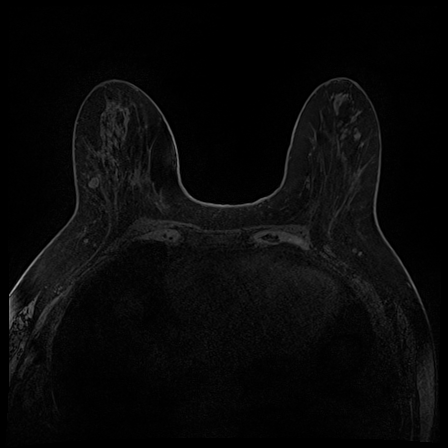}
    \includegraphics[height=1.4cm, width=.164\textwidth, trim={1.5cm 6.5cm 1.5cm 2.5cm},clip]{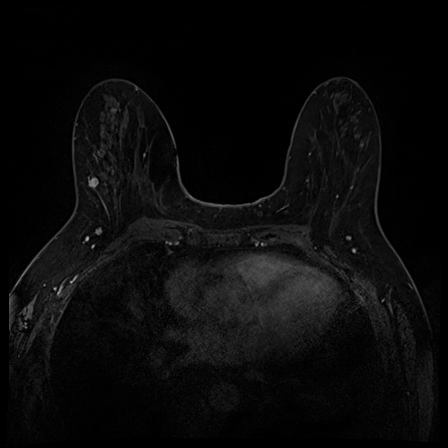}
    \includegraphics[height=1.4cm, width=.164\textwidth, trim={1.5cm 6.5cm 1.5cm 2.5cm},clip]{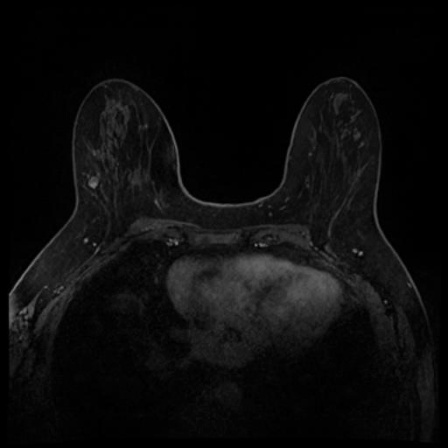}
    \includegraphics[height=1.4cm, width=.164\textwidth, trim={1.5cm 6.5cm 1.5cm 2.5cm},clip]{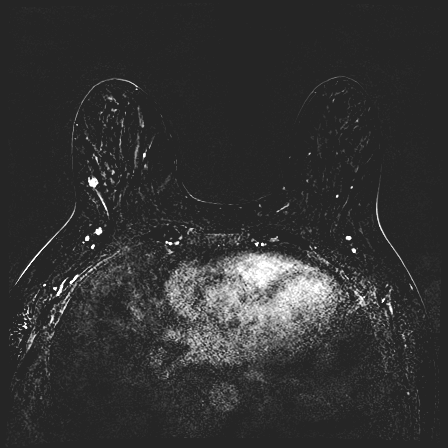}
    \includegraphics[height=1.4cm, width=.164\textwidth, trim={1.5cm 6.5cm 1.5cm 2.5cm},clip]{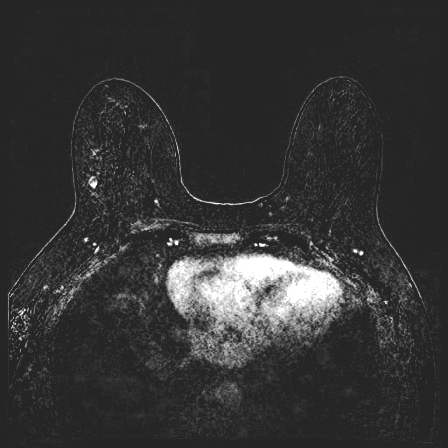}
    \includegraphics[height=1.4cm, width=.164\textwidth, trim={1.5cm 6.5cm 1.5cm 2.5cm},clip]{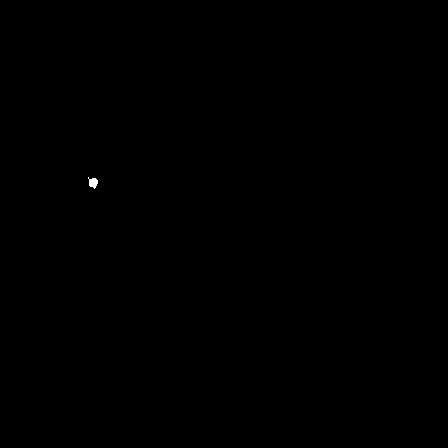}
    
    \includegraphics[height=1.4cm, width=.164\textwidth, trim={2.5cm 7.5cm 2.5cm 4cm},clip]{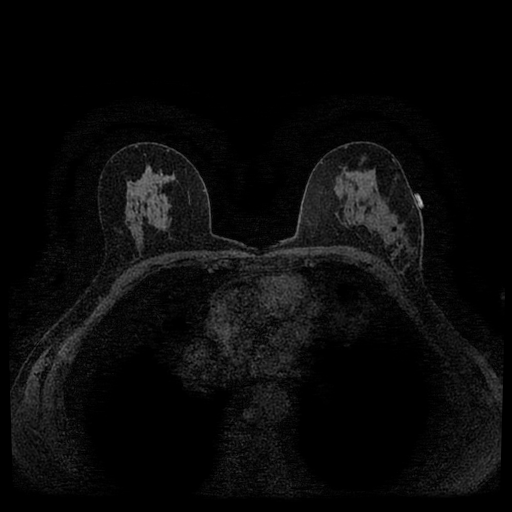}
    \includegraphics[height=1.4cm, width=.164\textwidth, trim={2.5cm 7.5cm 2.5cm 4cm},clip]{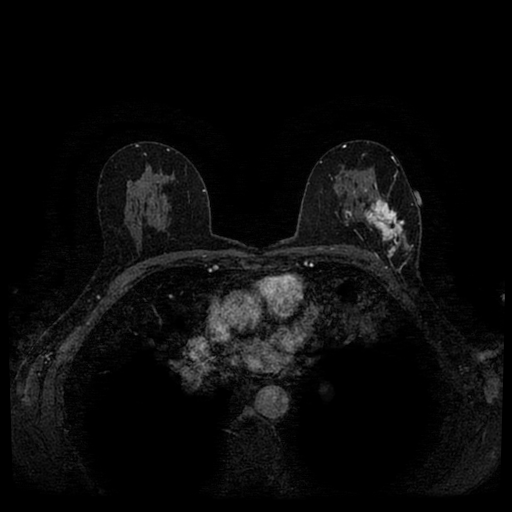}
    \includegraphics[height=1.4cm, width=.164\textwidth, trim={2.5cm 7.5cm 2.5cm 4cm},clip]{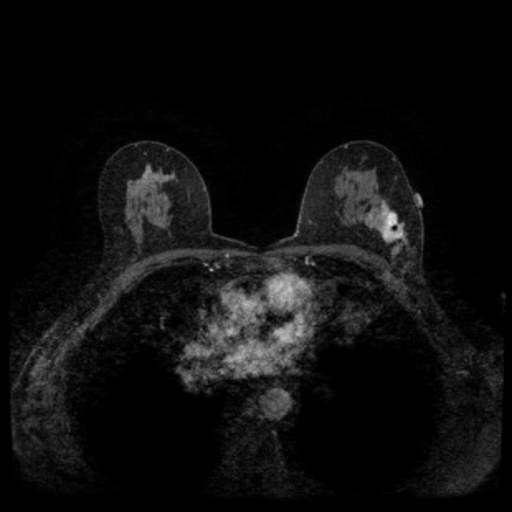}
    \includegraphics[height=1.4cm, width=.164\textwidth, trim={2.5cm 7.5cm 2.5cm 4cm},clip]{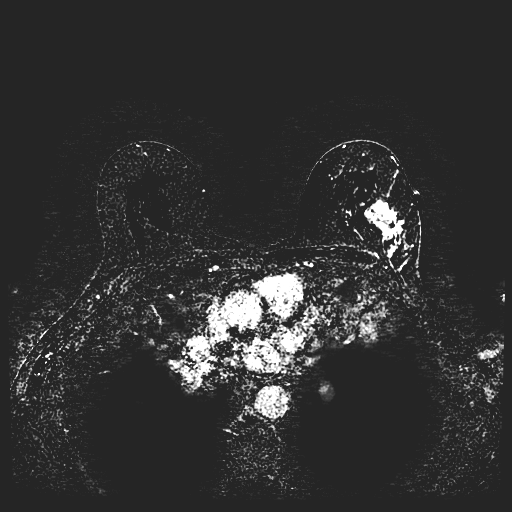}
    \includegraphics[height=1.4cm, width=.164\textwidth, trim={2.5cm 7.5cm 2.5cm 4cm},clip]{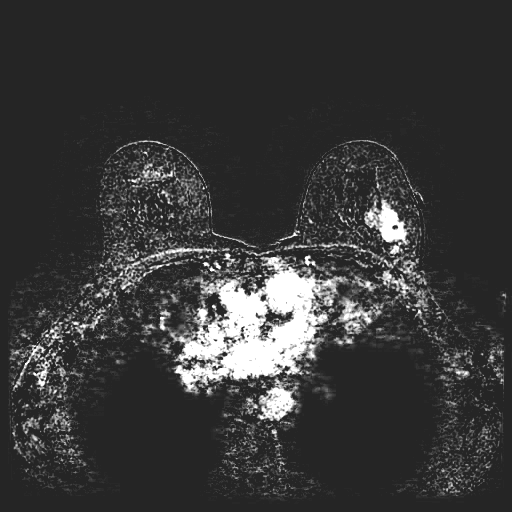}
    \includegraphics[height=1.4cm, width=.164\textwidth, trim={2.5cm 7.5cm 2.5cm 4cm},clip]{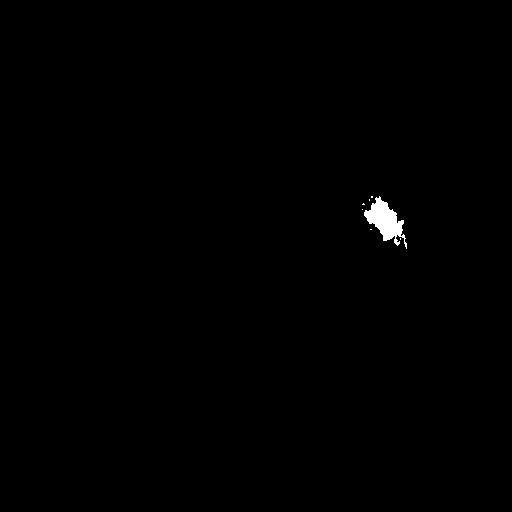}
    
    \includegraphics[height=1.4cm, width=.164\textwidth, trim={0 8.5cm 0 1.15cm},clip]{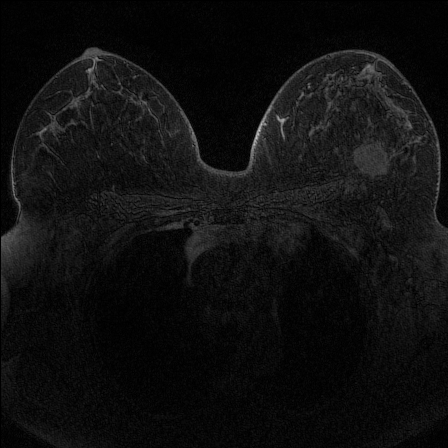}
    \includegraphics[height=1.4cm, width=.164\textwidth, trim={0 8.5cm 0 1.15cm},clip]{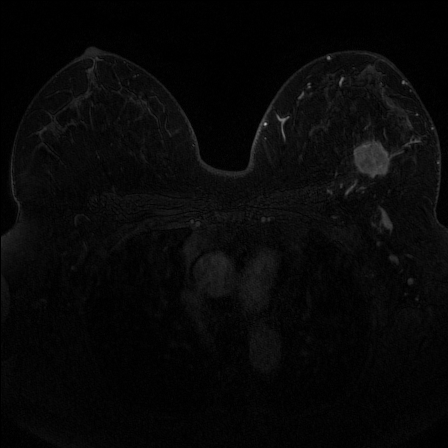}
    \includegraphics[height=1.4cm, width=.164\textwidth, trim={0 8.5cm 0 1.15cm},clip]{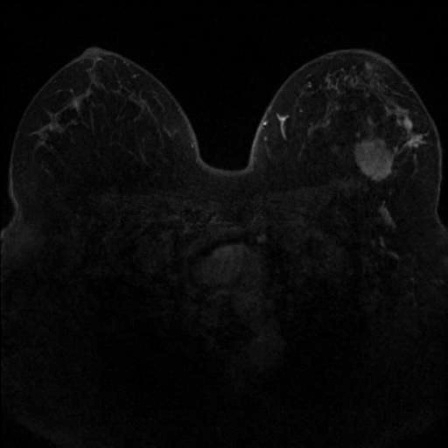}
    \includegraphics[height=1.4cm, width=.164\textwidth, trim={0 8.5cm 0 1.15cm},clip]{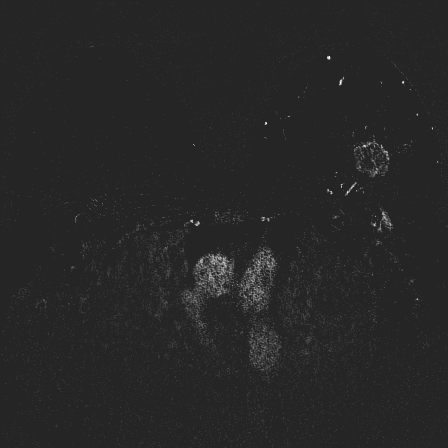}
    \includegraphics[height=1.4cm, width=.164\textwidth, trim={0 8.5cm 0 1.15cm},clip]{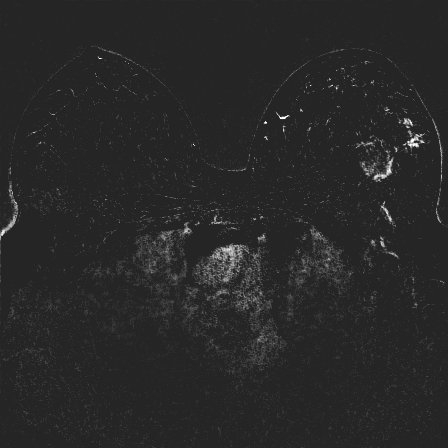}
    \includegraphics[height=1.4cm, width=.164\textwidth, trim={0 8.5cm 0 1.15cm},clip]{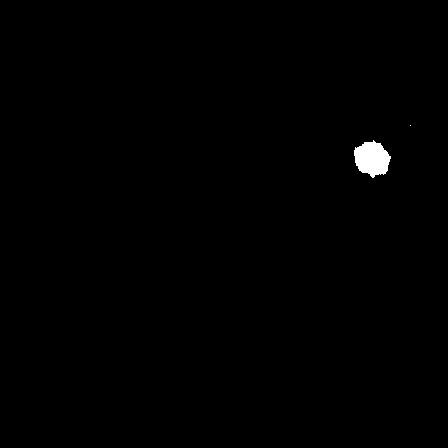}

    \includegraphics[height=1.4cm, width=.164\textwidth, trim={1.5cm 8.5cm 1.5cm 3cm},clip]
    {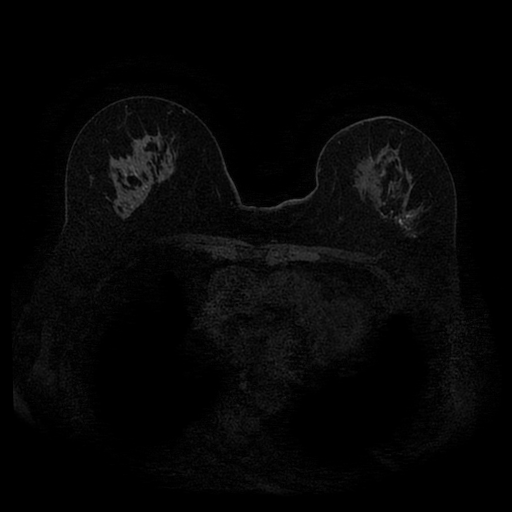}
    \includegraphics[height=1.4cm, width=.164\textwidth, trim={1.5cm 8.5cm 1.5cm 3cm},clip]
    {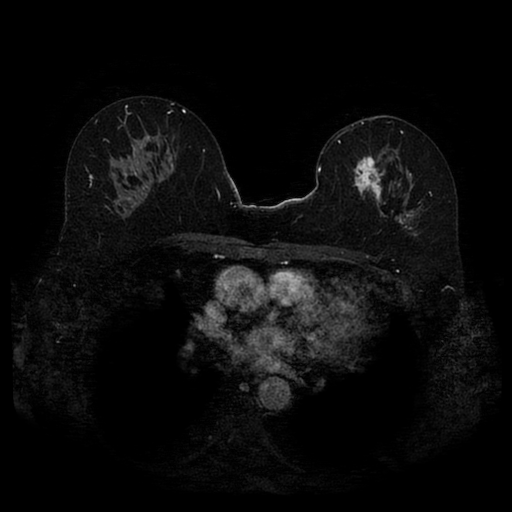}
    \includegraphics[height=1.4cm, width=.164\textwidth, trim={1.5cm 8.5cm 1.5cm 3cm},clip]
    {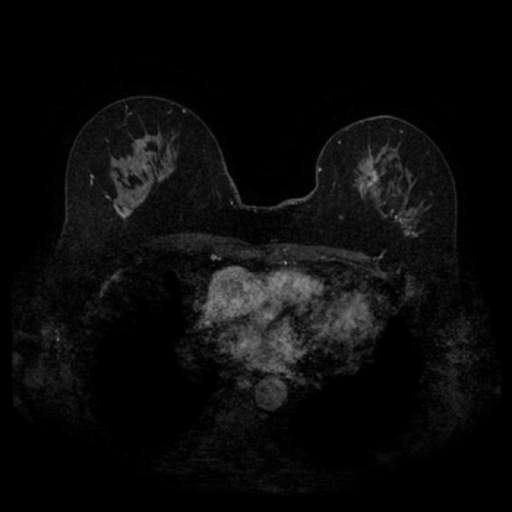}
    \includegraphics[height=1.4cm, width=.164\textwidth, trim={1.5cm 8.5cm 1.5cm 3cm},clip]{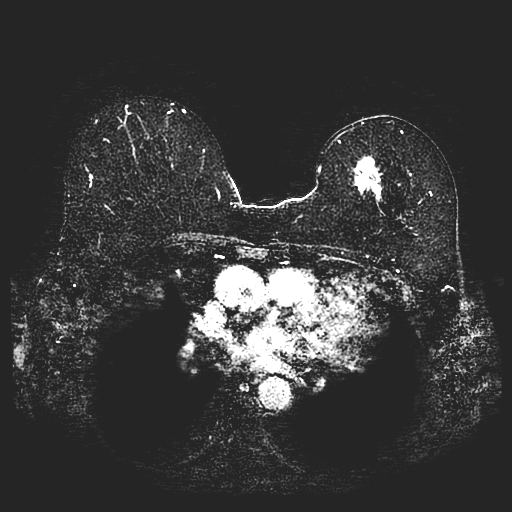}
    \includegraphics[height=1.4cm, width=.164\textwidth, trim={1.5cm 8.5cm 1.5cm 3cm},clip]{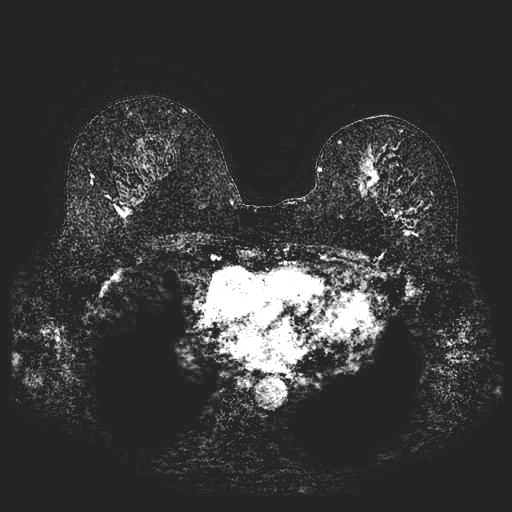}
    \includegraphics[height=1.4cm, width=.164\textwidth, trim={1.5cm 8.5cm 1.5cm 3cm},clip]
    {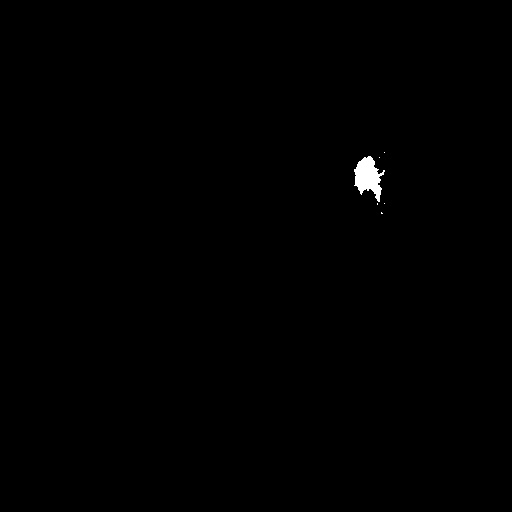}

    \includegraphics[height=1.4cm, width=.164\textwidth, trim={1.5cm 8.5cm 1.5cm 3.2cm},clip]{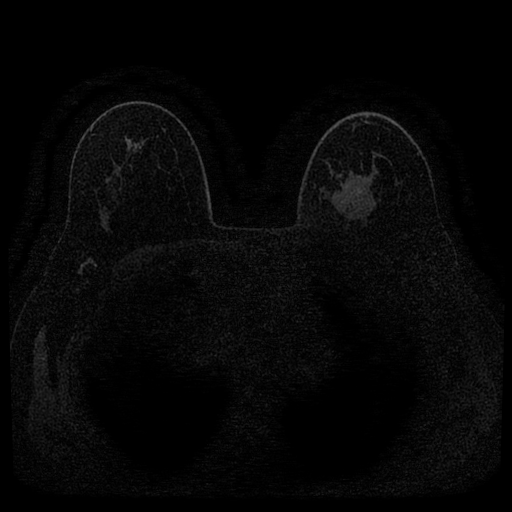}
    \includegraphics[height=1.4cm, width=.164\textwidth, trim={1.5cm 8.5cm 1.5cm 3.2cm},clip]{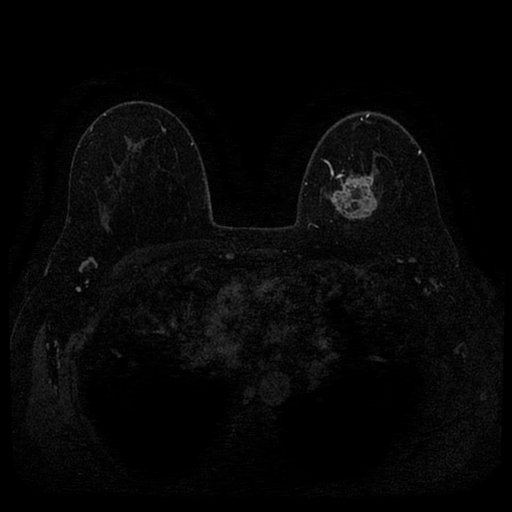}
    \includegraphics[height=1.4cm, width=.164\textwidth, trim={1.5cm 8.5cm 1.5cm 3.2cm},clip]
    {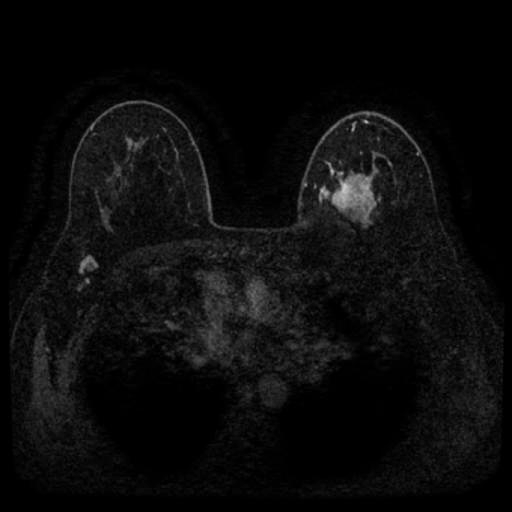}
    \includegraphics[height=1.4cm, width=.164\textwidth, trim={1.5cm 8.5cm 1.5cm 3.2cm},clip]{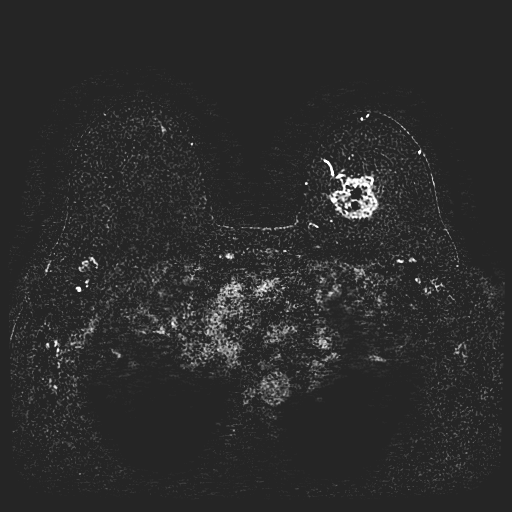}
    \includegraphics[height=1.4cm, width=.164\textwidth, trim={1.5cm 8.5cm 1.5cm 3.2cm},clip]{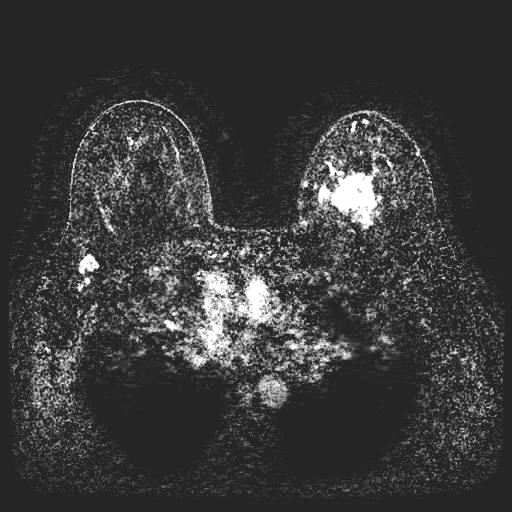}
    \includegraphics[height=1.4cm, width=.164\textwidth, trim={1.5cm 8.5cm 1.5cm 3.2cm},clip]
    {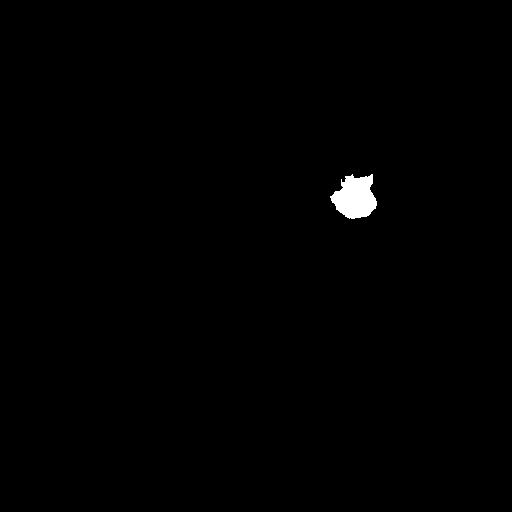}
    \caption{Synthesis of breast DCE-MRI as shown for six cases\cite{saha2018machine}. Two cases were manually selected from the validation set (\nth{1}
    row: Case 228 
    , 
    \nth{2} row: Case 886), two manually selected from the test set (\nth{3} row: Case 378, \nth{4} row: Case 907), and two randomly selected from the test set (\nth{5} row: Case 041, \nth{6} row: Case 045). From left to right, we illustrate axial slices of 
    the (\emph{a}) real T1-weighted pre-contrast MRI, (\emph{b}) the real DCE-MRI sequence 1, (\emph{c}) the synthetic DCE-MRI sequence 1, the subtraction image based on the (\emph{d}) real and (\emph{e}) synthetic DCE-MRI subtractions, and (\emph{f}) the ground truth segmentation mask. 
    }
 \label{fig:image_comparison}
 \end{figure}
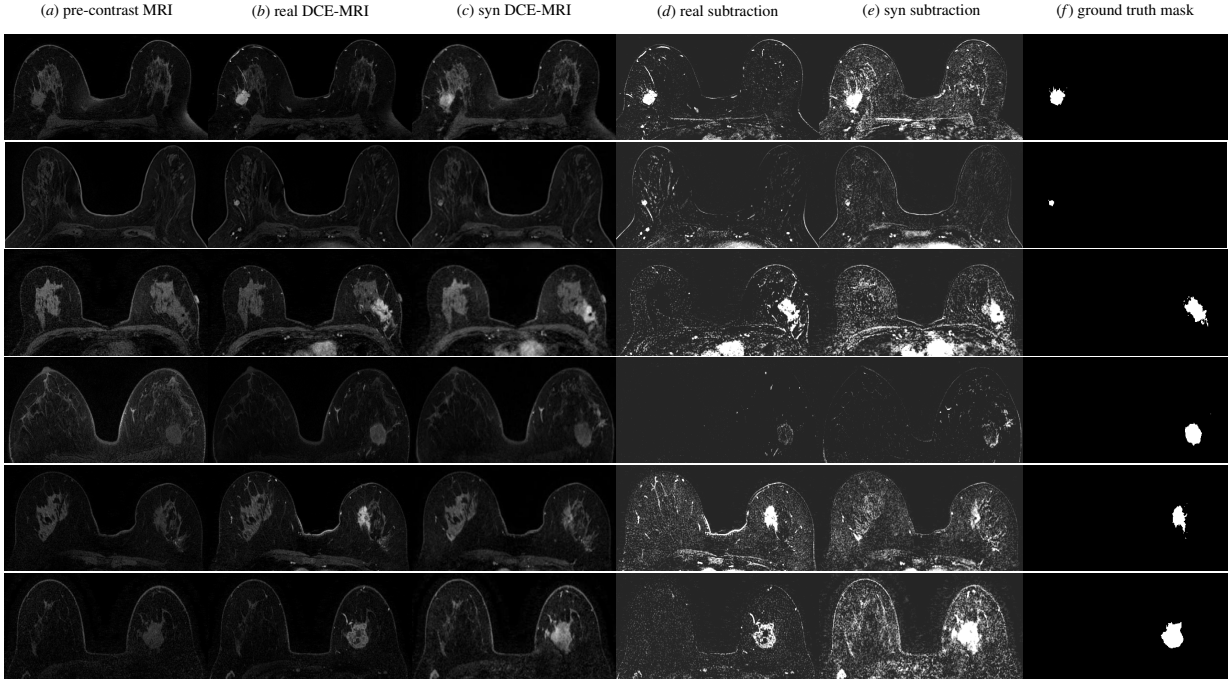

To systematically assess the quality of image synthesis, we compare metrics between synthetic and real post-contrast MRI slices in the test set. After training the generative model, we generate T1-weighted DCE-MRI phase 1 images - often corresponding to peak enhancement in the studied dataset\cite{saha2018machine} - for both the image synthesis test set (30 cases) and validation set (224 cases). Figures \ref{fig:image_comparison} present qualitative results, illustrating the model's translation of entire axial breast MRI slices to the post-contrast domain, along with corresponding subtraction images for six different patient cases. It is observed that some false-positive contrast regions are hallucinated (e.g., see the \nth{4} row), and some tumors are only partially contrast-enhanced (e.g., see the \nth{5} row of Figure \ref{fig:image_comparison}).
In the randomly chosen patient case 045, depicted in the \nth{6} row of Figure \ref{fig:image_comparison}, the real post-contrast image displays hypointense areas within the tumor, suggesting the presence of a necrotic core. \textcolor{mycorrect}{With this feature being not clearly visible} in the pre-contrast domain, it is not reproduced in the synthetic post-contrast image. \textcolor{mycorrect}{Nevertheless, the respective synthetic subtraction image enables detection and localization of the tumor, thereby preserving clinical utility for diagnostic and treatment workflows \cite{mann2019breast, saslow2007american} despite limitations in replicating the internal tumor microenvironment in full detail for contrast kinetics assessment.} Overall, the qualitative outcomes of our model underscore its capability to proficiently translate pre-contrast to DCE-MRI, demonstrating strong potential in synthetic contrast localization and enhancement.


%
\begin{table}[ht!]
\centering
\caption{Multi-metric synthetic image quality evaluation on the test set containing 5186 images. Synthetic images were generated after 30 GAN training epochs with SAMe score as epoch selection criterion. FID$_{Img}$ and FID$_{Rad}$ results are based on 2000 synthetic-real phase 1 DCE-MRI axial slice pairs, while 5000 pairs where used for the reminder of the metrics. \emph{Real DCE vs. Real Pre} describes the upper bound baseline that compares paired real pre-contrast and real phase 1 DCE slices. \emph{Subt} refers to subtraction images, where pre-contrast images are subtracted from either their real (\emph{Real Subt}) or synthetic (\emph{Syn Subt}) DCE counterparts. \emph{Splitted Test} describes a random by-patient split of the test set (i.e. without corresponding image pairs) that allows to capture the variance across patient cases in distribution comparsion metrics (i.e., FID).}
\vspace{2mm}
\resizebox{1.0\columnwidth}{!}{
\begin{tabular}{lcccccccccc}
    \toprule
     & & \multicolumn{9}{c}{Metric} \\
     \arrayrulecolor{light_grey} \cmidrule(lr){3-11}
    Comparison & Dataset &  FID$_{Img}$ $\downarrow$ & FID$_{Rad}$ & LPIPS $\downarrow$ & PSNR $\uparrow$ & SSIM $\uparrow$ & MS-SSIM $\uparrow$ & MAE $\downarrow$ & MSE $\downarrow$ \\
    \arrayrulecolor{black} \toprule
    Real DCE vs. Syn$_{ep30}$ DCE & Test &  \textbf{28.717} & \textbf{0.0385} & \textbf{0.064$\pm$0.04} & \textbf{32.91$\pm$1.35} & \textbf{0.726$\pm$0.089} & \textbf{0.798$\pm$0.08} & 85.623$\pm$38.297& \textbf{34.882$\pm$10.520} \\
    Real DCE vs. Real Pre  & Test & 59.644 & 0.1556 & 0.084$\pm$0.05& 32.42$\pm$1.68& 0.705$\pm$0.104& 0.780$\pm$0.07& \textbf{66.121$\pm$34.473} & 40.124$\pm$16.183\\
    \arrayrulecolor{light_grey} \cmidrule(lr){1-11}
    Real Subt vs. Syn$_{ep30}$ Subt & Test & 46.931 & 0.2864 & 0.062$\pm$0.03& 34.74$\pm$1.73& 0.692$\pm$0.097& 0.717$\pm$0.09& 44.896$\pm$23.403& 23.425$\pm$8.602\\
    \arrayrulecolor{light_grey} \cmidrule(lr){1-11}
    Real DCE vs. Syn$_{ep30}$ DCE & Splitted Test & \textbf{43.865} & 0.7012 & & & & & & & \\
    Real DCE vs. Real DCE              & Splitted Test  & 49.808  & \textbf{0.2060} & & & & & & & \\
    \arrayrulecolor{black} \bottomrule
\end{tabular}
}
\label{tab:image_quality_TEST}
\end{table}

Table \ref{tab:image_quality_TEST} presents a comparison of the 2D full axial slice image dataset, evaluating the similarity between the synthetic and real test case images. In this analysis, the synthetic DCE-MRI images demonstrate a significantly closer semantic and perceptual resemblance (as measured by FID scores, LPIPS, SSIM, MS-SSIM) to the real DCE-MRI phase 1 images compared to their real pre-contrast counterparts.
In the comparison of \emph{splitted test} datasets, it is important to note that the compared sets do not correspond to the same patient cases, enabling the assessment of variability across different patient test cases. Interestingly, based on the domain-agnostic FID$_{Img}$, the variability between real DCE-MRI cases is found to be higher than that between real and synthetic DCE-MRI cases. Conversely, the FID$_{Rad}$ indicates less variability between real DCE-MRI datasets than between real and synthetic ones for the same dataset split. Specifically, according to the radiology domain-specific FID$_{Rad}$, the variability across patient cases (\emph{splitted test}) is generally higher than the variability between the pre-, post-, and synthetic post-contrast sequences (\emph{test}) of corresponding cases. For real vs. synthetic DCE-MRI, this also holds for FID$_{Img}$. 

Additionally, we assess subtraction images created by subtracting a pre-contrast image from either its real or synthetic DCE-MRI counterpart. Compared to the real vs. synthetic DCE-MRI images, the corresponding real (\emph{Real Subt}) vs. synthetic (\emph{Syn}$_{ep30}$ \emph{Subt}) subtraction images show improved metrics in pretrained neural network-based image-level comparisons (i.e., LPIPS) and reconstruction-based metrics (i.e., MSE, MAE). However, at least this latter improvement can be attributed to the clipping of pixel values to 0 when they become negative after subtraction. In terms of structural perceptual metrics (i.e., SSIM, MS-SSIM) and latent feature distribution-based metrics (i.e., FID$_{Rad}$, FID$_{Img}$), the comparison between real and synthetic DCE-MRI images yields better quantitative results than the subtraction image comparison.

\subsection{Tumor Segmentation in First DCE-MRI Sequence}

Given the potential variability in data availability between pre- and post-contrast domains across different clinical environments, we conduct four types of tumor segmentation experiments. The first set of experiments, shown in the \nth{1} block of Table~\ref{tab:experiments_1_2}
, assumes that ground truth post-contrast data is entirely unavailable for segmentation during both training and testing. In this scenario, available pre-contrast training cases (\emph{baseline 1}) are augmented with their synthetic post-contrast counterparts. The second set of experiments assumes that pre-contrast data is available for training, while the test data consists of post-contrast images. Here, we evaluate tumor segmentation performance under a domain shift, examining the impact of adding synthetic post-contrast cases to the pre-contrast \emph{baseline 2}. The third set in Table~\ref{tab:experiments_1_2} considers a scenario where real post-contrast data is available and used for both training and testing, assessing whether synthetic data can enhance the performance of the post-contrast \emph{baseline 3}. Finally, the fourth set of experiments investigates a situation where segmentation models are trained on real post-contrast data but tested on pre-contrast cases. This scenario includes instances where contrast agents are not administered, such as in patient sub-populations like pregnant women, patients with kidney issues, those who decline contrast media, or those at high risk of allergic reactions to contrast agents.

\begin{table}[ht!]
\centering
\caption{
Tumor volume segmentation results across four scenarios: (1) pre-contrast domain only, (2) domain shift with post-contrast testing and no access to real post-contrast data during training, (3) combined pre- and post-contrast training with post-contrast testing, and (4) synthetic post-contrast data aiding models trained in the post-contrast domain but tested on pre-contrast data (e.g., due to patient pregnancy or allergy). Synthetic data improves performance notably in domain shift and pre-contrast test cases. Reported Dice coefficients are from a model ensemble with each model trained via one 5-fold cross-validation fold.
}
\vspace{2mm}
\resizebox{0.95\columnwidth}{!}{
\begin{tabular}{lc}
    \toprule
     Scenario \emph{1}. \small{Pre-contrast training data and pre-contrast test data available} &  Dice $\uparrow$ \\ \emph{train on:} & \emph{test on:} Real pre-contrast  \\
    \arrayrulecolor{light_grey} \toprule
    Real pre-contrast (\emph{baseline 1}) & \textbf{0.569}  \\
    Real pre-contrast + syn post-contrast (augmentation) & 0.531  \\
    %
    Syn post-contrast & 0.486 \\
    \arrayrulecolor{black} \toprule
     Scenario \emph{2}. \small{Domain shift: Pre-contrast training data, but no pre-contrast test data available}  &  \\ \emph{train on:} & \emph{test on:} Real post-contrast  \\
    \arrayrulecolor{light_grey} \toprule
    Real pre-contrast (\emph{baseline 2}) & 0.484  \\
    Real pre-contrast + syn post-contrast (augmentation) & 0.663  \\
    %
    Syn post-contrast & \textbf{0.687} \\
    \arrayrulecolor{black} \bottomrule
        Scenario \emph{3}. \small{Post-contrast training data and post-contrast test data available}  & Dice $\uparrow$ \\  \emph{train on:} & \emph{test on:} Real post-contrast  \\
    \arrayrulecolor{light_grey} \toprule
    Real post-contrast (\emph{baseline 3}) & 0.790 \\
    Real post-contrast + syn post-contrast (augmentation)  & \textbf{0.797} \\
    Real post-contrast + real pre-contrast (augmentation) &  0.780 \\
    Real post-contrast + real pre-contrast + syn post-contrast (augmentation) & 0.770 \\
    Syn post-contrast & 0.687 \\
    %
    \arrayrulecolor{black} \toprule
     Scenario \emph{4}. \small{Domain shift: Post-contrast training data, but no post-contrast test data available}  & \\ \emph{train on:} &  \emph{test on:} Real pre-contrast   \\
    \arrayrulecolor{light_grey} \toprule
    Real post-contrast (\emph{baseline 4}) & 0.164 \\
    Real post-contrast + syn post-contrast (augmentation)  & 0.409 \\
    Syn post-contrast & \textbf{0.486} \\
    \arrayrulecolor{black} \bottomrule
\end{tabular}
}
\label{tab:experiments_1_2}
\end{table}

In all data augmentation experiments, each training case is supplemented with its corresponding augmented version (e.g., real and/or synthetic post-contrast volumes). Importantly, the model does not receive any indication that an original training data point (e.g., a pre-contrast volume) and its augmented counterpart (e.g., a synthetic post-contrast volume) pertain to the same patient. The reported Dice coefficients are based on ensemble predictions from five segmentation models trained using 5-fold cross-validation\cite{isensee2021nnu}, which is why standard deviations are not reported.

Reviewing the results for \emph{baseline 1} in Table \ref{tab:experiments_1_2} , we observe that synthetic post-contrast augmentations do not enhance segmentation performance within the pre-contrast domain. However, in the domain shift context of \emph{baseline 2}, the inclusion of synthetic DCE-MRI volumes leads to a marked improvement in the post-contrast domain. Specifically, augmenting real pre-contrast data with synthetic post-contrast images increases the post-contrast Dice coefficient by $0.179$ (from $0.484$ to $0.663$), while maintaining a similar performance level in the pre-contrast domain (with Dice scores of $0.531$ compared to $0.569$). This finding aligns with the image quality analysis in Table \ref{tab:image_quality_TEST}, which confirms that the GAN-generated images fall within the post-contrast domain distribution, highlighting their effectiveness in addressing domain shift scenarios.
\emph{Baseline 3} demonstrates strong tumor segmentation performance in the post-contrast domain, achieving a Dice score of $0.790$. Although the improvement with synthetic DCE-MRI augmentation is modest, it still enhances performance to $0.797$, making it preferable over pre-contrast augmentations, which yield a slightly lower score of $0.780$. In contrast, \emph{baseline 4} shows a more significant Dice score increase of $0.245$ (from $0.164$ to $0.409$) in the pre-contrast test domain when synthetic post-contrast augmentations are used. Despite the synthetic DCE-MRI images being closely aligned with the DCE-MRI distribution (as indicated by an FID$_{Rad}$ of $0.0385$ between synthetic and real post-contrast test data in Table \ref{tab:image_quality_TEST}), it nonetheless captures relevant pre-contrast signals that enable the post-contrast segmentation model to generalize more effectively to pre-contrast test data. Notably, training solely on synthetic images, without real post-contrast counterparts, further boosts segmentation performance in the post-contrast domain by $0.077$ (from $0.409$ to $0.486$).

\subsection{Joint Synthesis of Multiple DCE-MRI Sequences}

\begin{table}[ht!]
\centering
\caption{Quantitative synthetic image quality assessment for joint synthesis of multiple DCE-MRI image timepoints \textcolor{mycorrect}{(Phase P1 to P3)}. Results include standard deviation where applicable and are based on 100 test cases consisting of 4422 image pairs. Best results in bold.
\emph{Real Post vs. Real Pre} indicates the lower bound baseline comparison between corresponding real pre-contrast MRI and DCE-MRI images. 
\textcolor{mycorrect}{GAN denotes our proposed Pix2PixHD-based post-contrast image synthesis method. GAN Subt refers to subtraction images created by subtracting real pre-contrast images from GAN-generated DCE-MRI images. U-Net refers to the subtraction image synthesis baseline method adapted and modified from \textit{Schreiter et al} \cite{schreiter2024virtual}, which directly predicts subtraction images from pre-contrast images.} 
\textcolor{mycorrect}{In FRD$_{BB+Tex}$} 
\textcolor{mycorrect}{, \textit{BB} denotes that feature extraction is conditioned by tumor bounding box annotations \cite{saha2018machine}, while \textit{Tex} denotes that only  texture features  were extracted \cite{van2017computational}, i.e., glcm, glrlm, gldm, glszm, ngtdm.} 
}
\resizebox{1.0\columnwidth}{!}{
\begin{tabular}{ll|ccccccccc}
    \toprule
    \multicolumn{2}{c}{512x512 breast MRI slices with tumor} & \multicolumn{9}{c}{Metrics} \\
     \arrayrulecolor{black} \cmidrule(lr){0-1} \cmidrule(lr){3-11}
    Set 1 & Set 2 & FID$_{Img}$ $\downarrow$ & FID$_{Rad}$ $\downarrow$ & \textcolor{mycorrect}{FRD$_{BB+Tex}$ $\downarrow$} &
     LPIPS $\downarrow$ & PSNR $\uparrow$ & SSIM $\uparrow$ & MS-SSIM $\uparrow$ &  MSE $\downarrow$ & MAE $\downarrow$ \\
    \arrayrulecolor{black} \toprule
    Real Pre-Contrast & Real DCE$_{P1}$ & 
    39.47 & 
    0.143 &
    209.70 & 
    .201 $\pm$ .098 & 
    31.88 $\pm$ 1.303 & 
    .712 $\pm$ .080 & 
    .785 $\pm$ .062 & 
    38.98 $\pm$ 11.88 & 
    109.55 $\pm$ 38.78 \\ 

    GAN: Syn DCE$_{P1}$ & Real DCE$_{P1}$ & 
    \textbf{20.32} &  
    \textbf{14.91} &
    \textbf{116.37} & 
    \textbf{.139 $\pm$ .071} & 
    \textbf{32.40 $\pm$ 1.153} & 
    \textbf{.749 $\pm$ .059} & 
    \textbf{.862 $\pm$ .042} & 
    \textbf{32.88 $\pm$ 8.331} & 
    \textbf{77.65 $\pm$ 37.50} \\

    \arrayrulecolor{light_grey} \cmidrule(lr){1-11}

    \textcolor{mycorrect}{
    U-Net: Syn Subt$_{P1}$} & \textcolor{mycorrect}{Real Subt$_{P1}$} & 
    296.65 & 
    14.91 &
    2046.91 & 
    .223 $\pm$ .713 & 
    10.67 $\pm$ 5.786 & 
    .620 $\pm$ .103 &
    .474 $\pm$ .124 & 
    \textbf{12.84 $\pm$ 7.113} & 
    \textbf{21.44 $\pm$ 9.704} \\
    
    \textcolor{mycorrect}{GAN: Syn Subt$_{P1}$} & \textcolor{mycorrect}{Real Subt$_{P1}$} & 
    \textbf{49.22} &  
    \textbf{4.531} &
    \textbf{264.25} & 
    \textbf{.175 $\pm$ .073} & 
    \textbf{23.63 $\pm$ 1.966} & 
    \textbf{.686 $\pm$ .081} & 
    \textbf{.690 $\pm$ .088} & 
    24.56 $\pm$ 9.031 & 
    51.84 $\pm$ 25.86 \\
    
    \arrayrulecolor{black} \bottomrule
    
    Real Pre-Contrast & Real DCE$_{P2}$ & 
    36.18 & 
    0.124 &  
    202.32 & 
    .182 $\pm$ .085 & 
    31.92 $\pm$ 1.383 & 
    .718 $\pm$ .082 &
    .811 $\pm$ .056 &
    38.80 $\pm$ 13.02 & 
    112.61 $\pm$ 41.55 \\
    
    GAN: Syn DCE$_{P2}$ & Real DCE$_{P2}$ & 
    \textbf{15.45} & 
    \textbf{0.062} &
    \textbf{98.19} & 
    \textbf{.131 $\pm$ .065} & 
    \textbf{32.59 $\pm$ 1.121} & 
    \textbf{.764 $\pm$ .059} &
    \textbf{.871 $\pm$ .041} & 
    \textbf{31.08 $\pm$ 7.686} & 
    \textbf{83.37 $\pm$ 38.74} \\

    \arrayrulecolor{light_grey} \cmidrule(lr){1-11}

    \textcolor{mycorrect}{U-Net: Syn Subt$_{P2}$} & \textcolor{mycorrect}{Real Subt$_{P2}$} & 
    282.00 & 
    14.07 &
    1986.08 & 
    .212 $\pm$ .069 & 
    11.56 $\pm$ 5.771 & 
    .630 $\pm$ .105 &
    .503 $\pm$ .131 & 
    \textbf{12.58 $\pm$ 6.729} & 
    \textbf{22.22 $\pm$ 10.02} \\
    
    \textcolor{mycorrect}{GAN: Syn Subt$_{P2}$} & \textcolor{mycorrect}{Real Subt$_{P2}$ } & 
    \textbf{27.88} &  
    \textbf{2.839} &
    \textbf{202.48} & 
    \textbf{.156 $\pm$ .066} & 
    \textbf{24.15 $\pm$ 1.99} & 
    \textbf{.698 $\pm$ .079} & 
    \textbf{.712 $\pm$ .091} & 
    24.25 $\pm$ 8.646 & 
    56.81 $\pm$ 25.00 \\

    \arrayrulecolor{black} \bottomrule
    Real Pre-Contrast & Real DCE$_{P3}$ &  
    33.79 & 
    0.121 &
    223.68 & 
    .175 $\pm$ .082 & 
    31.95 $\pm$ 1.383 & 
    .721 $\pm$ .081 &
    .821 $\pm$ .053 & 
    38.59 $\pm$ 12.95 & 
    112.67 $\pm$ 42.04 \\
    
    GAN: Syn DCE$_{P3}$ & Real DCE$_{P3}$ & 
    \textbf{14.38} & 
    \textbf{0.079} &
    \textbf{108.13} & 
    \textbf{.129 $\pm$ .066} & 
    \textbf{32.64 $\pm$ 1.122} & 
    \textbf{.765 $\pm$ .060} &
    \textbf{.871 $\pm$ .043} & 
    \textbf{30.75 $\pm$ 7.495} & 
    \textbf{80.88 $\pm$ 37.31} \\

    \arrayrulecolor{light_grey} \cmidrule(lr){1-11}
    
    \textcolor{mycorrect}{U-Net: Syn Subt$_{P3}$ } & \textcolor{mycorrect}{Real Subt$_{P3}$} & 
    260.21 & 
    8.17 &
    1679.39 & 
    .182 $\pm$ .062 & 
    12.91 $\pm$ 5.761 & 
    .638 $\pm$ .113 &
    .550 $\pm$ .133 & 
    \textbf{12.60 $\pm$ 5.684} & 
    \textbf{23.82 $\pm$ 11.64} \\
    
    \textcolor{mycorrect}{GAN: Syn Subt$_{P3}$ } & \textcolor{mycorrect}{Real Subt$_{P3}$} & 
    \textbf{49.71} &  
    \textbf{0.342} &
    \textbf{155.16} & 
    \textbf{.159 $\pm$ .067} & 
    \textbf{23.71 $\pm$ 2.067} & 
    \textbf{.683 $\pm$ .083} & 
    \textbf{.692 $\pm$ .094} & 
    25.04 $\pm$ 8.563 & 
    57.41 $\pm$ 23.98 \\

    \arrayrulecolor{black} \bottomrule
\end{tabular}
}
\label{tab:multi_timepoint}
\end{table}

Based on the respective pre-contrast T1-weighted image of each patient, we jointly generate the images corresponding to the first three DCE-MRI sequences acquisitions using a checkpoint after 30 training epochs of our multi-sequence conditional GAN. 

As shown in Table \ref{tab:multi_timepoint}, we \textcolor{mycorrect}{use} multiple metrics to quantitatively assess each generated DCE-MRI sequence based on its similarity to its respective real DCE-MRI sequence. To facilitate interpreting and evaluating the obtained metrics, we compute the metrics also for the similarity of pre-contrast images with images from each of the real DCE-MRI sequences. It is observable that across each of the three DCE-MRI sequences, the synthetically generated images \textcolor{mycorrect}{by our GAN method} are substantially closer to the real DCE-MRI images compared to the lower bound pre-contrast based baseline. This holds true across both image-distribution comparison metrics (FID$_{Img}$, FID$_{Rad}$) and image-level comparison metrics (e.g., LPIPS, SSIM, MSE). With the exception of MAE and FID$_{Rad}$ (DCE$_{P2}$ to DCE$_{P3}$), an improvement across metrics is observed with temporally later DCE-MRI acquisitions, indicating enhanced performance at subsequent imaging phases. The achieved  FID$_{Img}$ values (20.32 in DCE$_{P1}$) are notably low even when compared to the single DCE-MRI sequence experiments (28.71) from section \ref{sec:single-dce-mri-exp} indicating that the training and joint generation using multiple DCE-MRI sequences is likely to positively affect model performance.

\textcolor{mycorrect}{To quantitatively evaluate the texture of the tumor area on the synthesized images, we employ the Fréchet Radiomics Distance (FRD) \cite{osuala2024towards}, comparing radiomics texture feature distributions extracted from both the real and synthetic images. The FRD$_{BB+Tex}$ metric we compute restricts feature extraction to tumor regions defined by bounding box annotations \cite{saha2018machine} for the 2737 respective image-annotation pairs for each dataset. In FRD$_{BB+Tex}$ we include exclusively the texture-based features of interest, namely glcm, glrlm, gldm, glszm, and ngtdm derived using the PyRadiomics toolkit \cite{van2017computational}. The values of these features are z-score normalized across all features from both datasets (e.g. synthetic and real) and, for better interpretation, are scaled to the range common for FID as defined and recommended in \textit{Osuala et al} \cite{osuala2024towards}. Our adapted formulation emphasizes tumor-specific texture, which has been identified as a particularly challenging aspect of contrast-enhanced image synthesis due to the subtle and heterogeneous appearance of cancerous lesions and the tumor microenvironment. By focusing the analysis on this dimension, FRD$_{BB+Tex}$ provides a robust measure of how well the generative model captures clinically relevant tumor characteristics in synthetic images. For our GAN-generated synthetic DCE-MRI data we observe a substantially lower FRD$_{BB+Tex}$ scores compared to the real pre-contrast vs real DCE comparison baseline, e.g. 116.37 and 209.70, respectively, for the first DCE-MRI timepoint. This trend is present across all of the temporal DCE-MRI sequences, demonstrating that tumor areas within our GAN-generated images overall capture a range of meaningful texture features that are present in the real DCE-MRI tumor areas. The GAN-generated subtraction images, while resulting in higher FRD$_{BB+Tex}$ compared to its generated DCE-MRI counterparts, obtained vastly lower FRD$_{BB+Tex}$ compared to the U-Net baseline across all DCE sequences.}

\textcolor{mycorrect}{Inspired by the work of Schreiter et al. (2024) \cite{schreiter2024virtual}, which to our knowledge represents the closest approach to multi-sequence DCE breast MRI slice generation, we implemented a benchmark U-Net model \cite{ronneberger2015u} for comparative evaluation. While our implementation is based on the MCO-Net architecture proposed by Schreiter et al., it has been adapted to fit the specific requirements of our application.}
\textcolor{mycorrect}{Different to their setup, we use a batch size of 16 and apply a sigmoid activation function in the final layer instead of tanh, as our output values lie within the range [0,1]. Unlike the original work, which incorporates T2-weighted and multi-b-value diffusion-weighted images as additional inputs, we restrict our input to T1-weighted non-contrast-enhanced MRI due to the constraints of our dataset. The U-Net is trained to generate subtraction images, i.e., computed by subtracting the non-contrast T1-weighted input from each corresponding post-contrast DCE sequence, across all three DCE timepoints. While we also experimented with directly predicting post-contrast images using this U-Net approach, the subtraction-based output yielded better qualitative and quantitative performance and we report the latter in Table \ref{tab:multi_timepoint}. }

\textcolor{mycorrect}{Overall, the U-Net generated subtraction images (compared to their real subtraction image counterparts) show considerably lower performance (apart from MSE and MAE metrics) across all DCE-MRI timepoints in terms of image distribution metrics, as well as image-to-image comparison metrics, when compared to the GAN-based approaches. The GAN-based approaches consist of (a) the comparison of GAN-generated DCE-MRI images with their real counterparts and (b) the comparison of real subtraction images with synthetic subtraction, where the latter are computed based on real pre-contrast images subtracted from GAN-generated DCE-MRI images. Subtraction images, particularly those generated by the U-Net, show strong performance on pixel-wise metrics such as MAE and MSE. This is in part likely due to the high proportion of low-intensity pixels present in both real and synthetic subtraction images, especially when compared to full DCE-MRI slices. While the GAN-generated subtraction images otherwise have generally slightly lower performance values compared to the GAN-generated DCE-MRI images, they (i) capture the real subtraction image distribution reasonably well (e.g. see FID$_{Img}$ of 27.88 for sequence 2 in Table \ref{tab:multi_timepoint}) and (ii) also show desirable results on the image-to-image comparison metrics (e.g. see LPIPS across all temporal sequences in Table \ref{tab:multi_timepoint}).}

\begin{figure}
    \centering
    \includegraphics[height=4.5cm, width=16.4cm, clip]{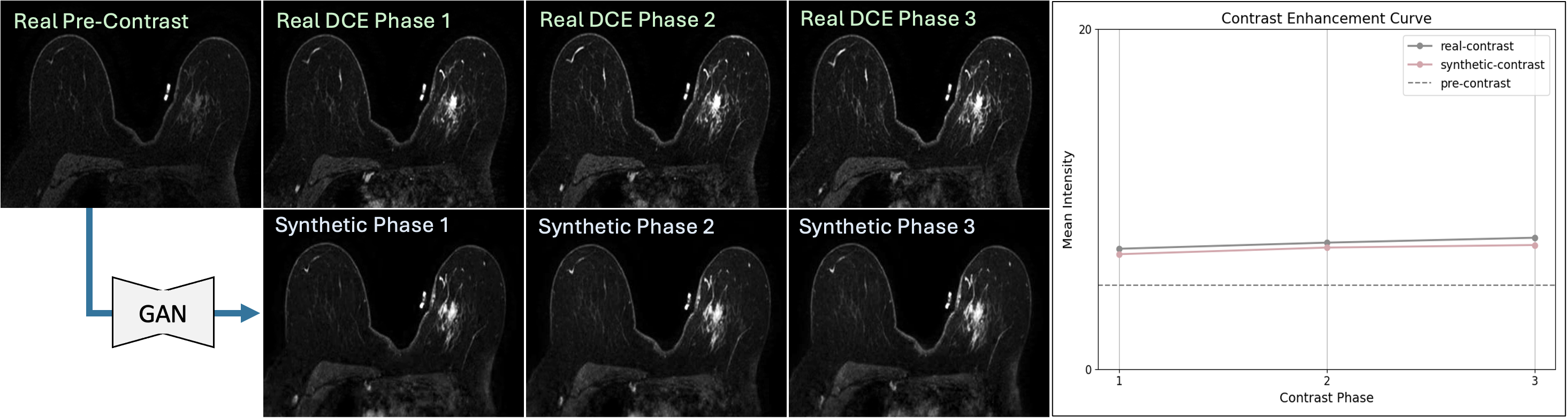}\hfill
    \includegraphics[height=4.5cm, width=16.4cm, clip]{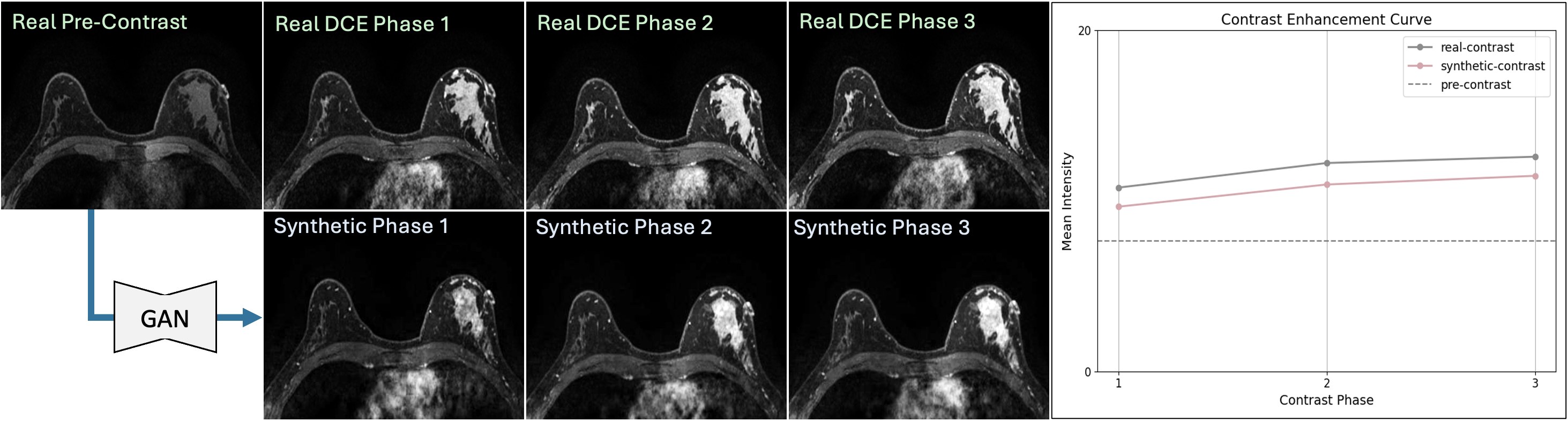}\hfill
    \includegraphics[height=4.5cm, width=16.4cm, clip]{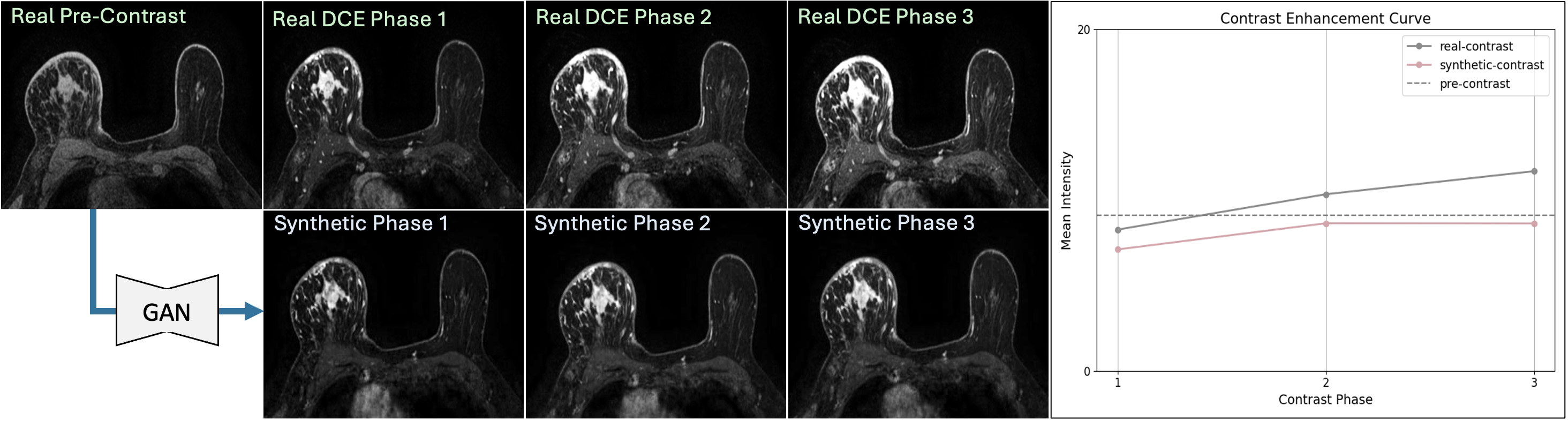}
    \caption{Qualitative results of joint multi-sequence DCE-MRI generation shown for 3 test cases with real images in each first row and respective synthetic images displayed in each second row. A single conditional GAN was trained to translate all test cases from pre-contrast to all three DCE-MRI sequences. \textcolor{mycorrect}{Contrast enhancement curves are also shown, visualizing changes in image intensity across consecutive temporal DCE-MRI sequences for both real and synthetic data.}}
 \label{fig:p1to3samples}
\end{figure}
Fig. \ref{fig:p1to3samples} provides a respective qualitative comparison of three patient cases across the first three DCE-MRI sequences. In these cases, and while noting a high similarity of images across DCE-MRI acquisitions, both in the real as well as in the synthetic images a trend of increased lesion contrast enhancement towards later DCE-MRI acquisitions is noticeable. \textcolor{mycorrect}{This temporal enhancement pattern is further reflected in the contrast enhancement curves, which show a generally increasing mean pixel intensity across the DCE-MRI timepoints. This trend is consistently observed in both the synthetic and real DCE data, indicating that the generative model captures patterns of the underlying contrast kinetics.}

\begin{figure}
    \centering
    \includegraphics[width=8cm]{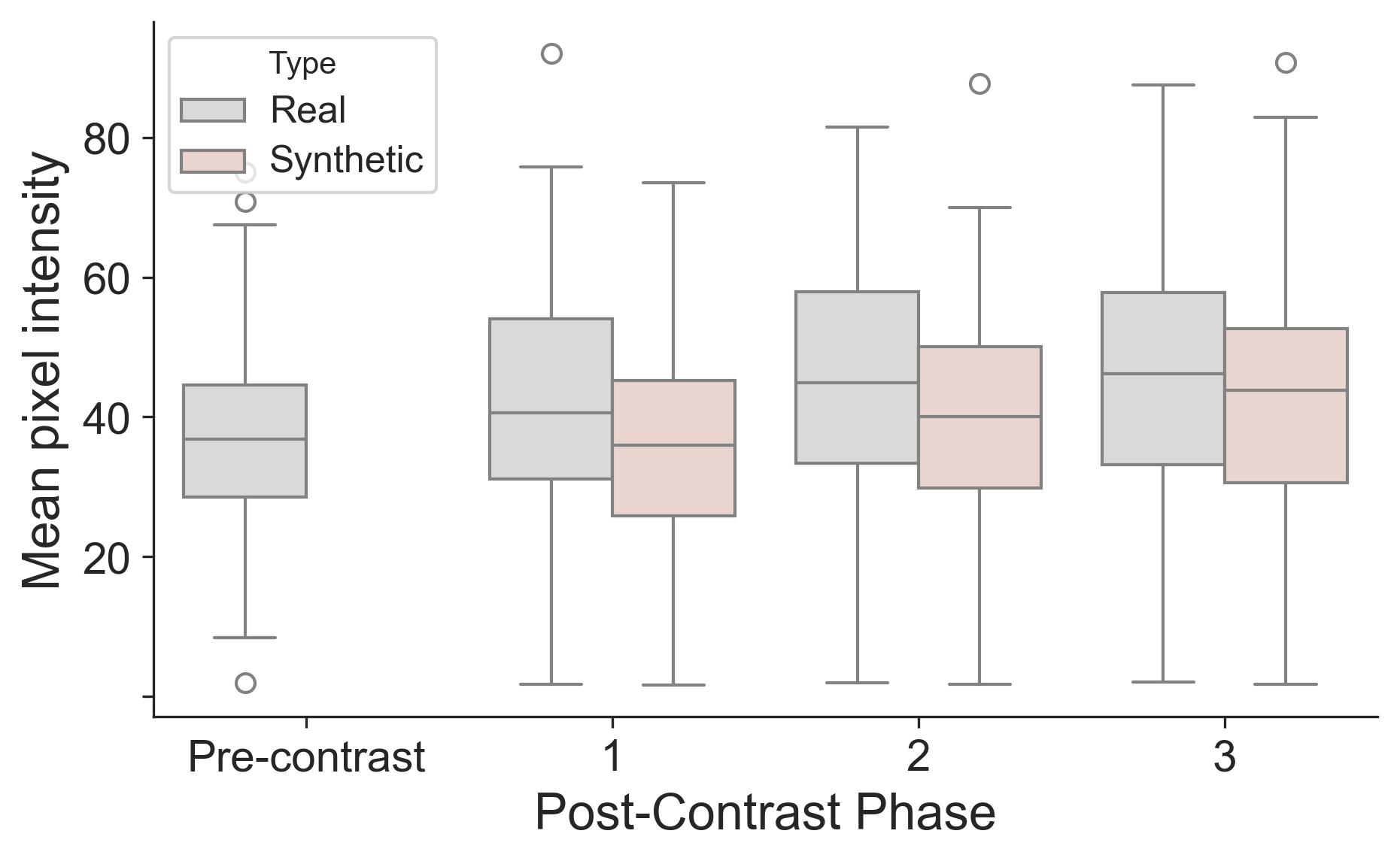}
    \caption{Illustration of temporal contrast enhancement patterns based on mean pixel intensity within the tumor bounding box area aggregated over all tumor-containing axial slices of all test cases. While gray coloring represents values computed for pre-contrast (T1) and post-contrast (DCE) of real test case images, red denotes values extracted from their synthetic DCE-MRI counterparts.}
 \label{fig:contrast-intensity-whole-set}
\end{figure}

To further explore the differences between DCE-MRI acquisitions, we additionally compute the mean and standard deviation of the pixel intensity within the tumor area, which we locate based on the bounding box information provided in the dataset \cite{saha2018machine}. The bounding box allows to capture and assess contrast uptake within the lesion but also, as opposed to lesion-level delineation, its closely surrounding tissue adjacent to the gross tumor region. Next, we aggregate the mean tumor intensity over all test cases to get a value and standard deviation for intensity for each of the three analyzed temporal DCE-MRI sequences. This process is repeated for pre-contrast and also for each synthetic DCE-MRI sequence, with the results being summarized and plotted in Fig. \ref{fig:contrast-intensity-whole-set}.

\begin{figure}
    \centering
    \includegraphics[height=7.6cm,width=15.3cm]{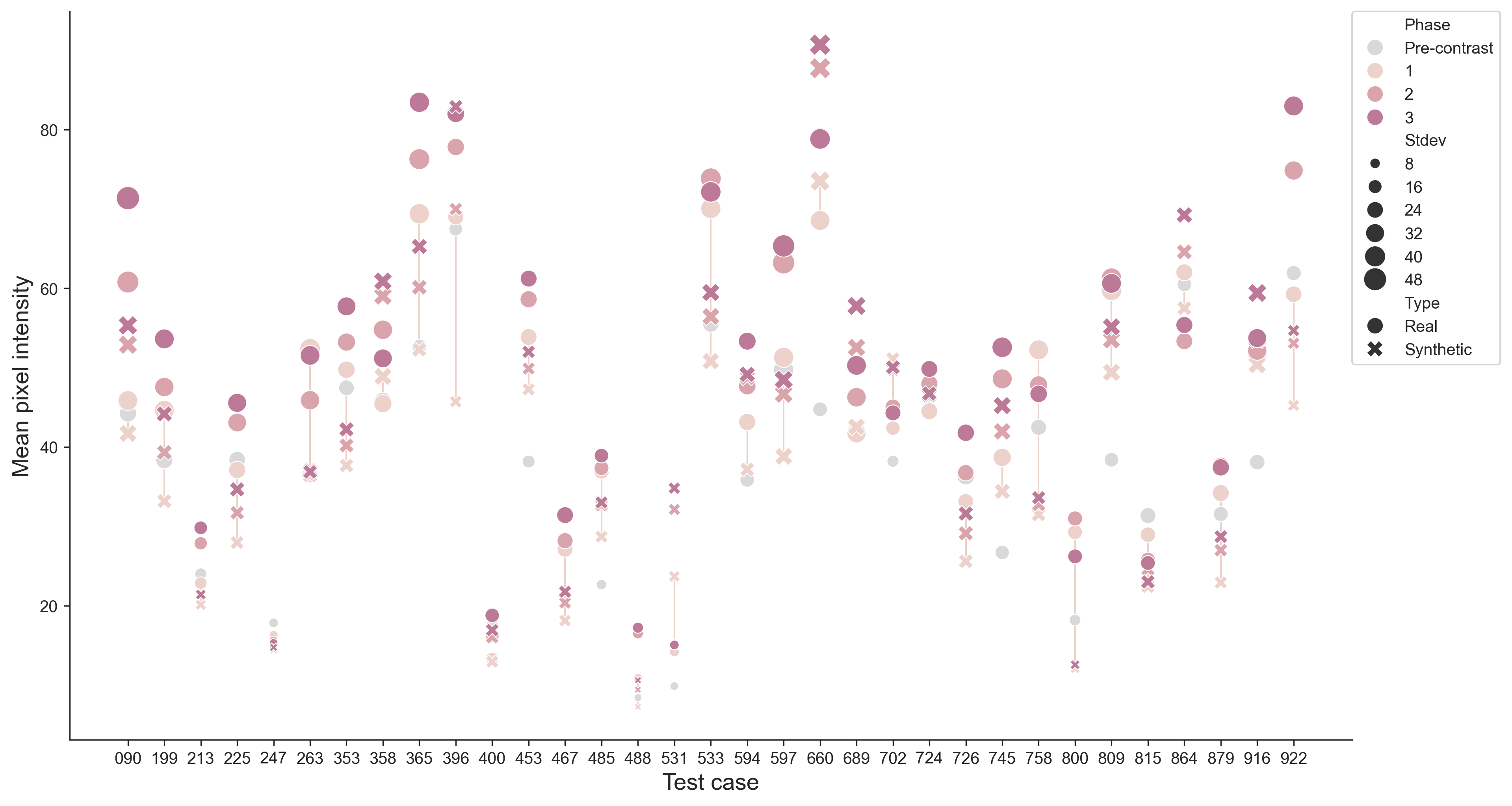}
    \caption{Visualization of temporal contrast enhancement patterns based on mean pixel intensity within the tumor bounding box area aggregated over all tumor-containing axial slices for each of a randomly selected 33 (out of 100) test cases. Circles denote real images while \textit{x} denotes synthetic counterparts. The mean pixel intensity standard deviation is represented by the marker size. The marker color encodes the temporal DCE-MRI sequence, with gray circles indicating the pre-contrast MRI sequence for comprehensiveness and comparability.}
 \label{fig:per-case}
\end{figure}

%
Focusing on the differences between real and synthetic post-contrast mean intensity temporal patterns, it is noted that the intensities generally increase towards later-stages of DCE-MRI acquisition, which is in line with aforedescribed visual assessment of Fig. \ref{fig:p1to3samples}. This trend in the real mean pixel intensity patterns is also followed by their synthetic counterparts. While the synthetic intensities have a slightly lower mean value than the real DCE-MRI ones, both synthetic and real intensities have comparable corresponding variances. In both cases, these variances are larger than the one of the pre-contrast tumor area intensity. This indicates more variety in the DCE-MRI domain, which is present in the synthetic DCE-MRI images. A rational for this diversity is given, for instance, by the inter- and intra-tumor heterogeneity often manifesting in a mixture of both hyper-intense and hypo-intense areas within the tumor in the DCE-MRI domain.
Moving from generative model evaluation towards clinical application, where contrast kinetics are used as biomarkers for tumor characterization, malignancy estimation, and treatment planning, we assess intensity changes per individual tumor area across DCE-MRI sequences. To this end, we randomly select 33 test cases and visualize their mean lesion intensity value for synthetic (denoted as x markers) and real (denoted as circle markers) for each sequence as well as for the respective pre-contrast image (denoted as gray circle marker) in Fig. \ref{fig:per-case}.

The observations in Fig. \ref{fig:per-case} corroborate conclusions drawn from Fig. \ref{fig:contrast-intensity-whole-set} indicating the general ability of the model to capture the dynamics of contrast uptake in the tumor lesion area.
In the more detailed per-case scheme in Figure \ref{fig:per-case}, the majority of the cases (22 out of 33), the synthetically generated images feature the right order for the mean pixel intensity values, increasing over time. Moreover, the standard deviation value between pixel intensity values in the tumor lesion bounding box, depicted by the marker size, can also be seen to be captured overall correctly by the generative model.
However, while synthetic data generally follows the trend of the real data, an off-set between the mean intensity values for the real and the synthetic images can be observed.
To this end, Fig. \ref{fig:per-case} demonstrates both, the complexity of the general task at hand, which asserts the generative model to detect, localize, highlight accordingly, and temporally adjust the highlighting for each heterogeneous tumor of each patient with variations in manufacturer, scanner, tumor molecular subtypes and patient characteristics. Overall, it is further indicated that the generative model is able to learn and represent this complexity reasonably well despite the differences between cases in contrast enhancement and tumor manifestation.

\section{Discussion and Future Work}
Based on the results described in Section \ref{sec:results}, we confirm our initial hypothesis that (post-contrast) DCE-MRI imaging data can be effectively generated from pre-contrast MRI inputs. Thus, the pre-contrast MRI provides a signal that is learnable via extracting statistical patterns from patient cohorts that allow the generative model to detect, localize, and enhance cancerous lesions. Our modeling approach can be viewed as unsupervised lesion detection method that enables to identify cancerous lesions without the need of respective lesion-level annotations. Our modeling paradigm is thus of particular interest for clinical settings considering the need for cost-effective deep-learning methods, where annotation conducted by clinical experts commonly represents a major cost-driving factor.
Towards clinical application it is to be further noted that synthetic data can occasionally produce false-positive tumor hallucinations. Despite of that, it can still be a valuable tool for localizing and highlighting potential anomalies within the MRI volume, as such anomalies can then be flagged for clinical revision and in-depth examination by clinical experts. In cases where imaging of both virtual contrast and injected contrast are available, the difference it is to be explored whether insights can be derived from the difference between the contrast enhancement in both images. For instance, if the virtual contrast prediction model is trained on only cancerous lesions (or e.g., tumors of a specific molecular subtype), then a different tumor manifestation in the real contrast enhancement can indicate an out-of-distribution case (e.g., benign, different molecular subtype). This consideration sheds light on the potential lying in research synthesizing multiple possible (contrast-enhanced) lesion manifestations to explore the within-distribution diversity and, thus, the uncertainty of a contrast prediction for a specific patient case. A contrast prediction marked as being uncertain (e.g. based on a variance threshold) can warrant the administration of contrast media in a clinical setting, where the risk-benefit tradeoff of physical CA injection is evaluated on a case-by-case basis.

Apart from treatment applications, future clinical validation studies are also encouraged to investigate generative modeling methods as diagnostic tools in DCE-MRI as screening modality in high-risk populations (e.g., with change in BRCA1 or BRCA2). Validating CA-free MRI screening regimes with synthetic DCE-MRI requires evaluation on additional dataset, where, apart from cancerous lesions, a variety of benign findings are present. As compared to the present dataset assembled from neoadjuvant treatment cohorts, tumors in the screening regime can differ being smaller in  size, earlier in stage, while also the potentially younger median age of patients can affect manifestations on the imaging data.

Going beyond unsupervised lesion detection, the prediction of contrast enhancement kinetics via multi-sequence DCE-MRI generation unlocks further additional clinical use-cases. As also indicated by our work, the progress in the field of computer vision, which rapidly expands the capabilities of deep generative models, narrows the gap towards wide-spread clinical application of such complex temporal modeling tasks. 
We note that multi-sequence DCE-MRI synthesis can potentially result in higher quality synthetic images than single-sequence synthesis, likely due to the generative model having to learn more nuanced patterns in the temporal synthesis task.
Considering our exploratory results for multi-sequence synthesis of DCE-MRI, such synthetic temporal contrast enhancement patterns can be considered a promising alternative for patients where CA injection is too risky. In this realm, biomarkers based on contrast kinetics are to be further investigated, particularly towards a comparison between real and synthetic biomarkers. This can enable future studies to define a benchmark for the research community that assesses the clinical meaningfulness of a generative model's produced synthetic biomarkers. Such a benchmark can further be extended by assessing the usefulness of real and synthetic biomarkers for specific clinical downstream tasks such as lesion malignancy prediction or treatment response prognosis. We note additional potential for enhancement of our approach by integrating time-since-event variables \cite{osuala2024towards} that condition the generative model to generate DCE-MRI images for specific moments in time e.g. based on milliseconds passed since pre-contrast acquisition or CA injection. This methodological extension introduces new possibilities of clinical application such as
synthetically covering timepoints of interest in DCE-MRI protocols that could not have been acquired, for instance due to the limitation of lengthy acquisition times of MRI scans. Multi-variable conditioning of the generative model can further allow counterfactual generation of DCE-MRI sequences, where for instance the malignancy level, tumor subtype, patient age, tumor staging, selected treatment, or other variables can be varied thereby resulting in insightful comparisons of alternative tumor manifestations.

\textcolor{mycorrect}{
While an exhaustive hyperparameter search was beyond the scope of this work due to computational constraints, future research is to explore alternative training schedules and optimizer configurations \cite{loshchilov2017decoupled, hazimeh2023mind, zhang2024adam}, enhancing the stability and generalization of our generative models. Notably, the observed performance degradation beyond epoch 30 suggests that more refined learning rate strategies can help mitigate overfitting and enable sustained training stability. In this context, a future direction is to quantify the impact of these strategies, such as cosine annealing learning rate schedules, which have been shown to facilitate smoother convergence and help avoid sharp minima \cite{loshchilov2016sgdr} and learning rate warm-up strategies \cite{kalra2024warmup}.}

Additional potential realms of research include the comparison of fine-tuning vs training from scratch of the generative model, assessing different GAN and generative model architectures such as denoising probabilistic diffusion models\cite{sohl2015deep, song2019generative, ho2020denoising}. Using larger-scale datasets will further invite for a stratified analysis as per tumor type, CA type, bolus volume, scanner type, field strength, and clinical center. \textcolor{mycorrect}{A further line of research is to investigate the effect of integrating additional 2D and 3D imaging data such as diffusion-weighted (DWI) MRI, T2-weighted MRI, non-fat-saturated images as well as subtraction images either as (additional) inputs or as outputs that can be useful to enhance image fidelity \cite{zhang2023synthesis,liebert2024impact,schreiter2024virtual}. In this regard, such input modalities provide complementary information to conditioning the generative model likely resulting in increased anatomical accuracy and reliability of lesion characteristics of the synthetic post-contrast sequences, particularly in complex cases. For instance, T2-weighted images can depict lesion morphology better as well as perifocal or prepectoral edema within the breast, while DWI captures higher signal intensity without relying on contrast agent administration \cite{mann2019breast}. However, including these modalities is associated with additional challenges, such as the need for accurate image registration across modalities to avoid the introduction of respective alignment errors or artifacts, as well as the limited availability of such MRI acquisitions in real-world settings.}

Another avenue to explore is to probe whether DCE-MRI synthesis quality can be enhanced by \textcolor{mycorrect}{(a)} iteratively enlarging the proportion of tumor-containing slices \textcolor{mycorrect}{or (b) by gradually increasing the patch size around the tumor region-of-interest \cite{fischer2024progressive}} during training.

\textcolor{mycorrect}{As exemplified by row 1 in Figure \ref{fig:image_comparison}, we observe the difficulty in distinguishing fibroglandular tissue from tumor enhancement in contrast enhancement synthesis. This distinction is inherently challenging, especially in dense breasts, where benign fibroglandular enhancement can mimic malignant uptake. To avoid such cases to lead to an increase in false positives future work can apply region-aware loss functions or incorporating annotated fibroglandular maps can improve tissue differentiation in complex breast compositions.} In general, adding and correctly weighting a reconstruction, perceptual, or adversarial loss for only the tumor area in DCE-MRI synthesis holds potential to increase fidelity, diversity, and usefulness of the synthetic data. Such approaches can potentially allow models to better capture particular those tumor features that lie in the long-tail of the distribution such as, for instance, tumors with particular necrotic cores. Future studies can include these tumor-focused modifications additionally in the evaluation methods extending SAMe and its components by adding and weighting tumor-area restrict computation of metrics for 2D slices as well as 3D volumes.

Based on tumor segmentation, we show that stacked synthetic DCE-MRI volumes can be useful in increasing the robustness of downstream task models. Enhancing segmentation model generalizability across imaging domains is particularly advantageous in DCE-MRI, especially for patient populations restricted to pre-contrast imaging due to CA administration contraindications such as allergic reactions, absence of consent, pregnancy, or compromised renal function. Additionally, in some cases, fat-saturated DCE-MRI sequence may closely mimic pre-contrast images due to small-sized tumors, low CA doses or rapid washout, emphasizing the necessity for models that maintain robust performance across both pre-contrast and post-contrast settings.

\textcolor{mycorrect}{Focusing on the use of synthetic DCE-MRI for training via augmentation and domain adaptation, another dimension worth exploring is testing models trained on real post-contrast images using synthetic post-contrast images at inference. While this may offer technical insights into the interchangeability of real and synthetic data, it remains a clinically open and critical question whether decision-making can rely on synthetic inputs, given the risk of artifacts or hallucinations \cite{osuala2022data, cohen2018distribution, tivnan2024hallucination}. The clearer clinical utility of synthetic data lies in enhancing model performance as training input, both within and across modalities, where it has shown value in improving generalization and robustness \cite{garrucho2023high, dorent2023crossmoda, campello2021multi, osuala2023medigan}.}

\textcolor{mycorrect}{As an alternative to data augmentation,} we guide future work to also explore pre-training on synthetic DCE-MRI for downstream model training. Related to that, a promising approach is to analyze the impact of concatenated corresponding multi-modality image inputs into the segmentation model during training and testing (e.g., pre-contrast with multiple DCE-MRI sequences, \textcolor{mycorrect}{as well as synthetic sequences}) while also \textcolor{mycorrect}{separately evaluating challenging bilateral and multifocal cases}.
\textcolor{mycorrect}{While this study explores synthetic DCE-MRI in the context of tumor segmentation, additional downstream applications such as radiomic feature analysis represent important directions for future research. Radiomic features quantify tumor heterogeneity, morphology, and enhancement dynamics, which are essential for diagnosis and treatment planning \cite{lambin2012radiomics}. In this context, the proposed SAMe metric may serve as a useful proxy for evaluating the preservation of structural and textural features relevant to radiomic tasks. Conversely, insights from radiomic analysis can inform future extensions of the SAMe metric to enhance its interpretability and robustness as a quantification tool of medical image synthesis. As a unified score of multiple quantitative measures, a next step in the evaluation of SAMe is to assess its alignment with clinical relevance. The latter can be assessed based on radiologist feedback or expert-driven image validation to correlate SAMe scores with visual assessments, diagnostic confidence, or lesion detectability. Guiding potential refinements to its weighting or formulation, such future work can help to establish the metric in clinical workflows.}

\section{Conclusion}
Following the SynTRUST framework\cite{osuala2022data} for trustworthy medical image synthesis studies, our work demonstrates that virtual contrast injection can generate high-quality synthetic DCE-MRI images, effectively supporting tumor detection, localization, and segmentation. Our findings highlight the potential of integrating deep generative models into MRI workflows as a non-invasive alternative for patients who cannot undergo standard contrast-enhanced imaging, thereby enabling more accurate and personalized treatment strategies.
The potential of DCE-MRI synthesis is further demonstrated as a training data augmentation method to enhance downstream breast tumor segmentation models, which, for instance, increases robustness across modalities. We further define trustworthy synthetic data based upon which we introduce the SAMe as a unified metric to evaluate synthetic data quality and guide generative model training checkpoint selection, addressing the limitations of conventional single-metric assessments. Additionally, generating multiple subsequent DCE-MRI sequences facilitates a deeper assessment of tumor response to contrast media, offering critical insights for tumor characterization and individualized care planning. Overall, this work marks a significant step toward incorporating virtual contrast and deep generative models into clinical practice, paving the way for improved diagnostic accuracy and patient outcomes in breast cancer management.

\subsection*{Disclosures}
The authors have no conflicts of interest to declare that are relevant to the content of this article.

\subsection* {Acknowledgements}
This project has received funding from the European Union’s Horizon Europe and Horizon 2020 research and innovation programme under grant agreement No 101057699 (RadioVal) and No 952103 (EuCanImage), respectively. Also, this work was partially supported by the project FUTURE-ES (PID2021-126724OB-I00) and AIMED (PID2023-146786OB-I00) from the Ministry of Science, Innovation and Universities of Spain. We would like to thank Dr Marco Caballo (Radboud University Medical Center, The Netherlands) for providing 254 segmentation masks for the Duke Dataset. RO acknowledges a research stay grant from the Helmholtz Information and Data Science Academy (HIDA). \textcolor{mycorrect}{Daniel M. Lang and Julia A. Schnabel received funding from HELMHOLTZ IMAGING, a platform of the Helmholtz Information \& Data Science Incubator.}

\subsection* {Data, Materials, and Code Availability} 
The data\cite{saha2018machine} used in this study is publicly available on \href{https://doi.org/10.7937/TCIA.e3sv-re93}{https://doi.org/10.7937/TCIA.e3sv-re93}. The code is available in a Github repository on \href{https://github.com/RichardObi/SimulatingDCE}{https://github.com/RichardObi/SimulatingDCE}. The trained models are readily usable in the \textit{medigan} library on \href{https://github.com/RichardObi/medigan}{https://github.com/RichardObi/medigan}.


\bibliography{report}   
\bibliographystyle{spiejour}   



\vspace{1ex}
\noindent Biographies and photographs of the other authors are not available.

\listoffigures
\listoftables

\end{spacing}

\end{document}